\documentclass[sigconf,natbib=true]{acmart}

\usepackage{multirow}
\newcommand\sys{\textsc{LexBoost}}

\usepackage{algorithm}
\usepackage{algpseudocode}
\usepackage{todonotes}
\usepackage{float}
\newfloat{algorithm}{t}{lop}
\newcommand{\pageenlarge}[1]{\enlargethispage{#1\baselineskip}}

\AtBeginDocument{%
  \providecommand\BibTeX{{%
    \normalfont B\kern-0.5em{\scshape i\kern-0.25em b}\kern-0.8em\TeX}}}

\copyrightyear{2024}
\acmYear{2024}
\setcopyright{acmlicensed}\acmConference[DocEng '24]{ACM Symposium on Document Engineering 2024}{August 20--23, 2024}{San Jose, CA, USA}
\acmBooktitle{ACM Symposium on Document Engineering 2024 (DocEng '24), August 20--23, 2024, San Jose, CA, USA}
\acmDOI{10.1145/3685650.3685658}
\acmISBN{979-8-4007-1169-5/24/08}

\begin{document}

\title{LexBoost: \\ Improving Lexical Document Retrieval with Nearest Neighbors}

\author{Hrishikesh Kulkarni}
\affiliation{%
  \institution{Georgetown University}
  \city{Washington, DC}
  \country{USA}}
\email{first@ir.cs.georgetown.edu}

\author{Nazli Goharian}
\affiliation{%
 \institution{Georgetown University}
 \city{Washington, DC}
 \country{USA}}
\email{first@ir.cs.georgetown.edu}

\author{Ophir Frieder}
\affiliation{%
 \institution{Georgetown University}
 \city{Washington, DC}
 \country{USA}}
\email{first@ir.cs.georgetown.edu}

\author{Sean MacAvaney}
\affiliation{%
  \institution{University of Glasgow}
  \city{Glasgow}
  \country{UK}
}
\email{first.last@glasgow.ac.uk}
\renewcommand{\shortauthors}{Hrishikesh Kulkarni, Nazli Goharian,  Ophir Frieder \& Sean MacAvaney}
\begin{abstract}
%by understanding its neighborhood 
Sparse retrieval methods like BM25 are based on lexical overlap, focusing on the surface form of the terms that appear in the query and the document. The use of inverted indices in these methods leads to high retrieval efficiency. On the other hand, dense retrieval methods are based on learned dense vectors and, consequently, are effective but comparatively slow. Since sparse and dense methods approach problems differently and use complementary relevance signals, approximation methods were proposed to balance effectiveness and efficiency. For efficiency, approximation methods like HNSW are frequently used to approximate exhaustive dense retrieval. However, approximation techniques still exhibit considerably higher latency than sparse approaches. We propose \sys{} that first builds a network of dense neighbors (a corpus graph) using a dense retrieval approach while indexing. Then, during retrieval, we consider both a document's lexical relevance scores and its neighbors' scores to rank the documents. In \sys{} this remarkably simple application of the Cluster Hypothesis contributes to stronger ranking effectiveness while contributing little computational overhead (since the corpus graph is constructed offline). The method is robust across the number of neighbors considered, various fusion parameters for determining the scores, and different dataset construction methods.
% Importantly, \sys{} leads to statistically significant improvements with virtually no additional latency overhead, since the corpus graph is computed fully offline.
We also show that re-ranking on top of \sys{} outperforms traditional dense re-ranking and leads to results comparable with higher-latency exhaustive dense retrieval. 
\end{abstract}






\maketitle
          
\section{Introduction}

%Information Retrieval (IR) is about locating and retrieving the relevant documents with reference to user information needs represented in the form of queries. Information retrieval embraces the intellectual aspects of information and its specification for search, and also whatever systems, techniques, or machines are employed to carry out the operation \cite{mooers}.
%Thus, Information Retrieval systems identify documents relevant to user queries using a wide range of lexical, neural approaches and their combinations. In this case, appropriateness for the user's needs and preferences drives the effectiveness while the timely retrieval decides efficiency of retrieval \cite{formoso}. Both, effectiveness and efficiency are  crucial as users demand the best possible results in the shortest possible time. 

\looseness -1 Traditionally, lexical relevance methods like BM25 \cite{Robertson2009ThePR} are used for efficient retrieval. These methods consider the surface form of query and document terms and operate using term overlap, ranking documents with more occurrences of important terms ahead of others. Since the surface form of terms are the same regardless of their context within the document, lexical methods tend to miss highly relevant documents due to the vocabulary mismatch problem, limiting their effectiveness in both precision and recall. 
To overcome this limitation, dense methods were proposed; they go beyond word overlap by learning semantic representations of the documents. 
%This approach matches queries and documents in a low-dimension embedding space utilizing deep neural networks to learn the low-dimensional representations \cite{Xiong2020ApproximateNN}. 
These methods utilize deep neural networks to learn the low-dimensional representations and match queries and documents in the low-dimension embedding space \cite{Xiong2020ApproximateNN}.
Dense and neural retrieval methods represent documents in vector space of predefined size. They successfully identify semantic closeness between the query and document, significantly boosting retrieval effectiveness.

\begin{figure*}
\centering
% \captionsetup{justification=centering}

\includegraphics[scale=0.40]{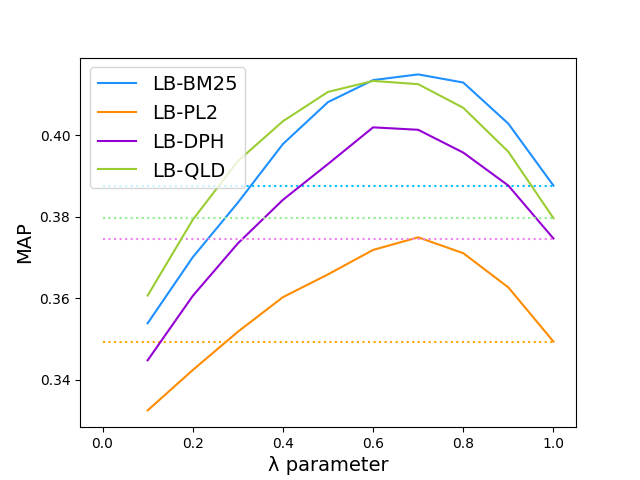}
\includegraphics[scale=0.40]{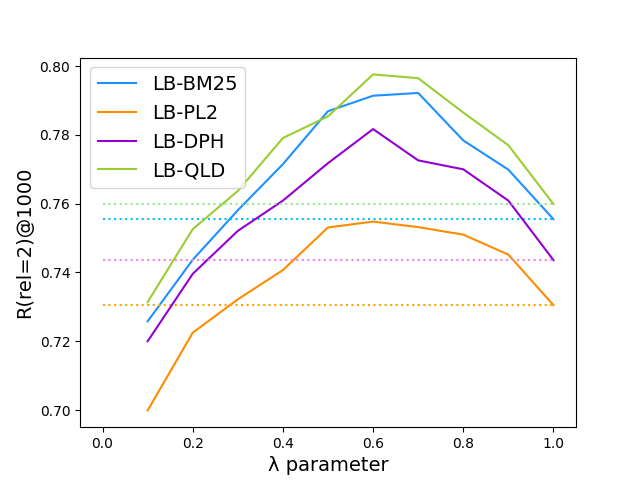}

\caption{MAP and Recall(rel=2)@1000 for \sys{} on BM25, PL2, DPH, QLD - TREC DL 2019. The faint horizontal lines are respective baselines (i.e., $\lambda=1$).}
\label{fig:teaser}
\end{figure*}

% ***** OF:  deleted next paragraph - unnecessary  ***

%Given the lexical and semantic nature of sparse and dense retrieval, they look at the problem from two different perspectives. Hence, the signals generated by them are complementary.  Hence, further improvements were achieved in hybrid methods which combined these complementary signals obtained from sparse and dense retrieval approaches. But, both dense as well as hybrid retrieval need an exhaustive search over all document vectors. This results in high latency costs along with exhaustive GPU use.
An ideal information retrieval system would 
thus capitalize on the
efficiency of lexical retrieval methods and the effectiveness of dense retrieval methods. The efficiency limitation of dense retrieval methods has led to a number of approximation approaches %by different researchers focusing on reduction in dense retrieval quantum through approximation. They 
that efficiently `approximate' the results of an exhaustive dense retrieval search, %more efficiently by proposing methods like re-ranking, 
including
HNSW \cite{8594636}, IVF \cite{1238663,DBLP:conf/eccv/YuanGCLJ12} and LADR \cite{10.1145/3539618.3591715}. These methods come close to a dense retriever and do so relatively efficiently. However, using a dense retrieval component still comes with substantially higher latency than lexical retrievers, thus only partially addressing the original problem of simultaneously achieving high effectiveness and efficiency.
 
We propose \sys{} that goes beyond lexical overlap by utilizing a document's neighborhood in dense retrieval space. Rather than directly combining sparse and dense scores, our approach estimates relevance by combining a document's lexical score with its neighbors' lexical scores. This is an application of the Cluster Hypothesis~\cite{Jardine1971TheUO}: we use a dense model to identify document proximity offline and a sparse model to estimate the relevance of a document (and its neighbors) online.
% Our novel approach of capturing the semantic relevance through a document's proximity to other documents and the use of these scores in a lexical paradigm helps our method to outperform popular lexical methods.
%\todo{all other state-of-the-art methods is a bit too broad. Maybe "perform competitively with other approaches, especially for the low time-budget"} 
%Further, \sys{} optimizes the inputs received through neighborhood to understand the relevance of every document in a better way.
%\todo{this sentence is unclear}

Figure \ref{fig:teaser} presents the improvements of \sys{} when used on top of lexical retrieval methods like BM25 \cite{Robertson2009ThePR}, PL2 \cite{10.1145/582415.582416}, DPH \cite{10.1145/582415.582416} and QLD \cite{qld}. 
%The flat lines are the baseline lexical retrieval results which are clearly outperformed by \sys{} Retrieval for a wide range of fusion parameter values.
Since the neighborhood is identified offline (i.e., at the indexing time), there is essentially no additional query-time latency overhead when using \sys{} as compared to existing lexical retrievers.
The enriched relevance information by effectively using knowledge about the neighborhood and negligible additional latency overheads separates \sys{} from other retrieval approaches. In summary, our contributions are:
\vspace{-0.5em}
\begin{itemize}
    \item We introduce \sys{}, which results in statistically significant improvements over the state-of-the-art lexical retrieval methods, including BM25.
    \item Our proposed method \sys{} introduces negligible additional latency overheads in achieving these statistically significant improvements.
    \item We conducted extensive experimentation using standard benchmark datasets in multiple domains and across a wide range of parameters and demonstrated improvements.
%    \item \sys{} is robust over the number of neighbors under consideration, use of different fusion parameters and datasets from different domains irrespective of their construction methods.
\end{itemize} 

\section{Related Work}

%With increasing volume of data resulting due to information explosion, more effective and efficient information retrieval systems are the need of time. 
Information retrieval is the process of matching the query against information items and obtaining the most relevant piece of information. This process involves creation of an index which is an optimized data structure built on top of the information items for faster access \cite{irbook}. A model (retrieval strategy) is an algorithm along with pertinent data structures which assigns similarity score to every query-document pair \cite{irbook}. Further, document-document similarity has also been explored for various applications \cite{10.1145/2644866.2644895}. Some of the foundational models of information retrieval are Boolean model - built on binary relevance, Vector Space model - uses spatial distance as a similarity measure and Probabilistic model - estimates document relevance as a probability \cite{liu2007web}. These models have heavily inspired the traditional language/vocabulary driven i.e. lexical information retrieval.

\subsection{Lexical Methods}

Traditionally, lexical methods based on word overlap were used \cite{Mikolov2013EfficientEO}. Utilization of inverted index in this sparse retrieval leads to high efficiency due to the term to document mapping. Different methods of weighing and normalization \cite{10.5555/1861751.1861756} led to a range of Term Frequency Inverse Document Frequency (TFIDF) models. BM25 \cite{Robertson2009ThePR} is one of the most popular and effective formulation in sparse retrieval. BM25 is a bag-of-words retrieval function. Here the documents are ranked based on the query terms appearing in each document, regardless of their proximity within the document. BM25 can be viewed as a non-linear combination of three basic document attributes: term frequency, document frequency, and the length of the document \cite{10.1145/1645953.1646237}. 
Document length normalization and query term saturation are the key features in BM25 and hence it did not favor shorter or longer documents and mitigates the impact of excessively high term frequency unlike TFIDF.
BM25 is also extended with prescription regarding how to combine additional fields in document description \cite{10.1145/1031171.1031181}.  
%Further, it computes BM25 across the expanded document description.

Divergence From Randomness (DFR) framework was proposed to build probabilistic term weighting schemes \cite{10.1145/582415.582416}. It consists of two divergence functions and one normalization function. Two of the best DFR framework models are PL2 (Poisson-Laplace with second normalization of term frequency) and DPH (hyper-geometric model Popper's normalization) \cite{10.1145/582415.582416}.
These methods typically suffer from high vocabulary dependence \cite{mitra2018an}. 
Query Linear Combination and Relevant Documents (QLD) uses relevant documents of similar queries for expressing the query as a linear combination of existing queries \cite{qld}.
Query expansion and pseudo relevance feedback offer resolve to some extent but come with certain latency overhead \cite{10.1145/2071389.2071390}. 
KL expansion \cite{10.1145/502585.502654}, Rocchio \cite{rocchio71relevance}, Relevance Modelling \cite{Metzler2005AMR} and RM3 \cite{abduljaleel2004umass} are popular pseudo relevance feedback methods.
Efficiency of lexical methods is exploited by using it as a first-stage document ranker for a short listed input to intricate dense retrieval and large language model based systems \cite{10.1145/3573128.3609340}.

\subsection{Dense Methods}

Further, to improve the effectiveness of information retrieval systems neural-based approaches for semantic retrieval, such as those utilizing CNNs and RNNs have been proposed. Here text documents and queries are represented in a continuous vector space. As a next step neural network based similarity measures are used for relevance calculation \cite{10.1145/2567948.2577348,10.1145/2505515.2505665}.
Additionally, neural methods utilize learned non-linear representations of text data, resulting in significant retrieval performance improvements \cite{10.1145/2567948.2577348}. Further, BERT \cite{devlin-etal-2019-bert}, GPT \cite{NEURIPS2020_1457c0d6} and other transformer architectures \cite{NIPS2017_3f5ee243},  have been used to improve the ability of information retrieval systems to attend to important parts of the query and documents for matching. 

Thus, dense retrieval methods focus on learning dense representations for documents and work on semantic level. This boosts the effectiveness by a great extent but at the same time an exhaustive search over all document vectors results in compromized efficiency.
Mainly, two primary types of dense methods exist: interaction-based where interactions between words in queries are modelled and the other being representation-based where the model learns a single vector representation of the query \cite{https://doi.org/10.48550/arxiv.1606.04648}. TAS-B \cite{10.1145/3404835.3462891} and TCT-ColBERT-HNP \cite{lin-etal-2021-batch} are some of the state-of-the-art result producing methods belonging to this category.
Further, multi representations method exemplified by ColBERT require significant memory and pruning methods have been proposed to make it more efficient \cite{10.1145/3573128.3604896}. Also, better sampling strategies have been proposed to train more effective dense retrieval models \cite{10.1007/978-3-031-56063-7_16}. Dense retrieval approaches have also been modified to perform entity-oriented document retrieval \cite{10.1007/978-3-031-56027-9_13}.

The advent of dense retrieval led to learned vector based pseudo relevance feedback models. Some of the popular ones include ColBERT PRF \cite{10.1145/3572405}, ColBERT-TCT PRF \cite{lin-etal-2021-batch} and ANCE PRF \cite{10.1145/3459637.3482124}. Models like CEQE (Contextualized Embeddings for Query Expansion) utilize query focused contextualized embedding vectors \cite{10.1007/978-3-030-72113-8_31}. Even though term based and vector based pseudo relevance feedback help tackle vocabulary dependence to some extent, they come with latency overheads. This is where \sys{} differentiates from others. 

\subsection{Approximation Methods}

Approximation methods approximate results of exhaustive dense retrieval efficiently for the purpose of enhanced efficiency. Even though they manage to approximate top results they lack in recall \cite{10.1145/3539618.3591715}.
Re-ranking is the most popular approximation method where lexical method like BM25 is used to shortlist top $n$ documents which are then re-ranked using costly dense retrieval methods. Here, the recall limitation is evident and its severity is determined by shortlisting parameter $n$. 
Other approximation methods can be classified into tree-based indexing, locality sensitive hashing and product-quantization-based and graph-based methods \cite{10.1145/3539618.3591651}. Popular approximation methods include HNSW \cite{8594636}, IVF \cite{1238663,DBLP:conf/eccv/YuanGCLJ12}, GAR \cite{macavaney:cikm2022-adaptive}, LADR \cite{10.1145/3539618.3591715} etc. These methods approximate results of a full dense retriever relatively efficiently. But the presence of a dense retriever component still adds additional latency overhead. 

\subsubsection{Comparison with LADR}

\sys{} uses corpus graph to re-rank the documents on the basis of scores assigned to their neighbors in the first-stage lexical retrieval. 
LADR uses corpus graph to identify additional potentially relevant documents to be re-ranked along with first-stage retrieval results by a dense retrieval method in the re-ranking stage.
Hence, \sys{} and LADR both use a corpus graph but for different purposes and at different stages in the multi-stage retrieval pipeline.
On the MS MARCO TREC DL 19 dataset \cite{https://doi.org/10.48550/arxiv.2003.07820}, LADR results in MAP of 0.50 using seed documents from BM25. In the same setup, it results in MAP of 0.51 when using seed documents from the BM25->\sys{} pipeline. This shows that \sys{} also results in improved effectiveness when used in first-stage retrieval for approximation methods like LADR at virtually no additional latency.

%These methods approximate results of a full dense retriever relatively efficiently. But the presence of a dense retriever component still adds additional latency overhead. LADR uses a corpus graph with the objective of identifying probably relevant documents in the second-stage of re-ranking using dense method. While, \sys{} uses corpus graph for assigning final scores to the documents in first-stage lexical retrieval.     

\begin{figure}
    \centering
\includegraphics[width=0.8\columnwidth]{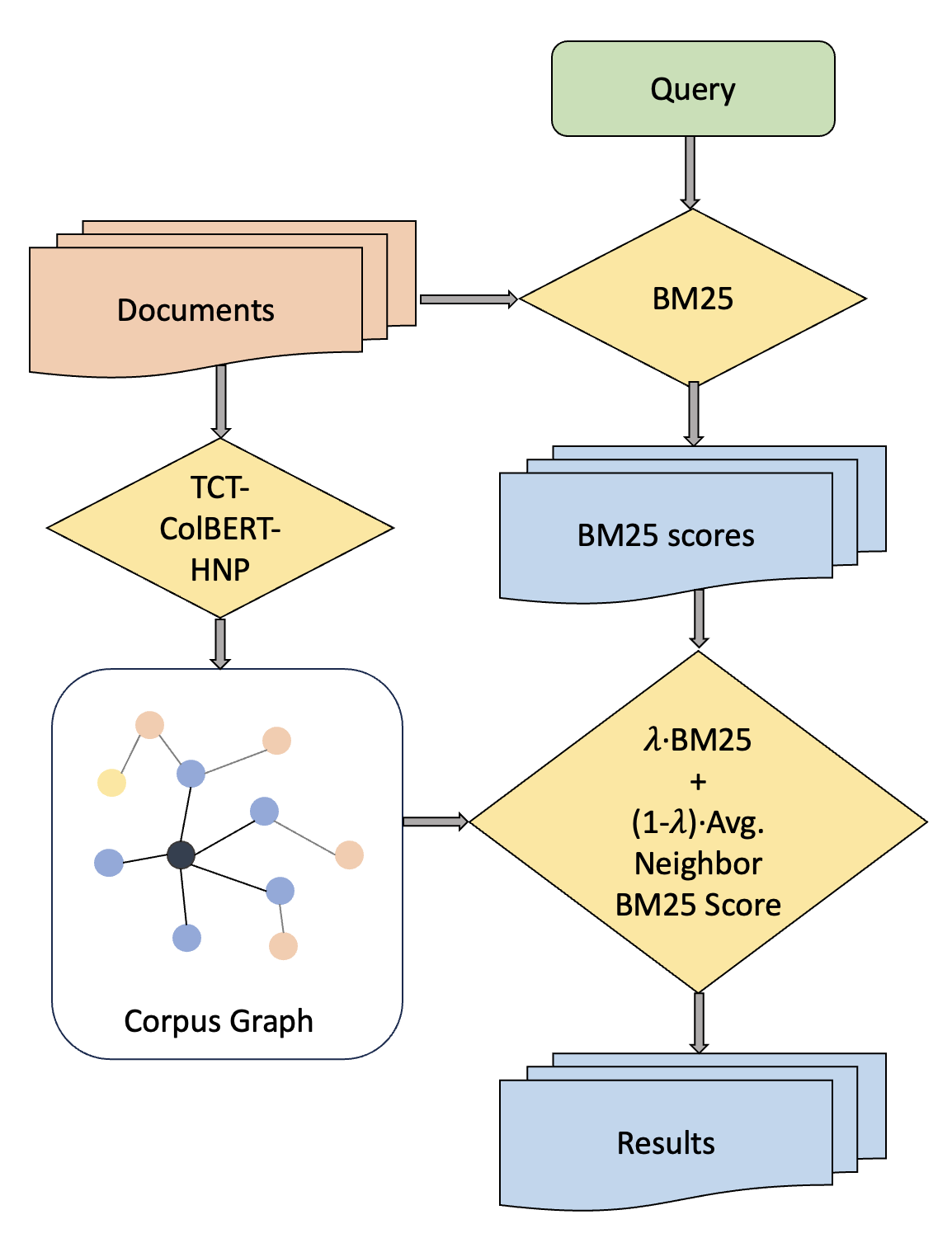}
    \caption{System Architecture}
    \label{fig:sysarch}
\end{figure}

\subsection{Hybrid Methods}
As sparse and dense retrieval methods provide complementary results - hybrid methods were proposed \cite{10.1145/3596512,NEJI20211111}. 
In any hybrid method, the lexical method and dense method output individual scores and ranked lists which are combined using various fusion formulations. Convex Combination of lexical and semantic scores and Reciprocal Rank Fusion of individual ranked lists are two of the most popular methods to combine output individual scores and ranked lists \cite{10.1145/3596512}. Convex Combination is more robust, agnostic to the choice of score normalization, has only one parameter and outperforms Reciprocal Rank Fusion \cite{10.1145/3596512}. Further, Convex Combination is also sample efficient with only a small set of examples required to tune the fusion parameter for the target domain \cite{10.1145/3596512}.

These hybrid methods have a dense retrieval component and hence have very high latency. Thus, the effectiveness comes with latency overhead and across the approaches in literature high latency remains an important issue to be looked at.
The key difference in our proposed method \sys{} is that the statistically significant improvement is delivered without invoking dense method at query time.

\section{\sys{}}

In \sys{}, we determine the final score of each document based on the lexical score (e.g., BM25) of the document and the lexical scores of its neighbors. The intuition behind the approach is derived from the Cluster Hypothesis~\cite{Jardine1971TheUO}, which states that documents that are near one another are likely to be relevant to the same query. Consequently, we consider a document to be more relevant if its nearest neighbors are also considered relevant.
% Here, we utilize insights from the location of the document in the corpus graph.
% \sys{} Retrieval derives and discovers knowledge from the individual's location along with the neighbors of the document.
Since the network of neighbors (i.e., the corpus graph) can be constructed offline, the online costs of applying dense retrieval techniques are mitigated.

Thus, the insights come from dense retrieval method based similarity but without any latency overhead during retrieval. Hence we name our proposed approach \sys{}. We use Convex Combination for fusion of the two scores given its effectiveness \cite{10.1145/3596512}. We define a $\lambda \in [0,1]$ parameter for fusion of the lexical method score and dense method based insight from the corpus graph. $\lambda$ parameter is tuned using a validation set for optimal results.

\begin{algorithm}
\caption{\sys{}}\label{alg:fh}
\begin{algorithmic}
\Require $q$ query, $D$ document corpus, $\lambda$ fusion parameter, $n$ number of neighbors, $G$ Corpus Graph
\State $R \gets \{\}$
\State $S \gets \Call{LexicalRetrieval}{q,D}$ \Comment{retrieve from corpus}
\For{d in S}
        % \State $s \gets \Call{ScoreLookup}{d, S}$ \Comment{look up doc score}
        \State $N \gets \Call{Neighbors}{d, \text{top }n\text{ from }G}$ \Comment{get neighbors}
        % \State $T \gets \Call{ScoreLookup}{N, S}$ \Comment{lookup neighbor scores}\\
        \State $s \gets \lambda\cdot \Call{Lookup}{d,S} + \dfrac{1-\lambda}{n}\cdot\underset{d_{neigh}\in N}{\sum}\Call{Lookup}{d_{neigh}, S}$\\
        % \Comment{get \sys{} score}\\
        \State $R[d] \gets s$ \Comment{save \sys{} doc score}
        
      \EndFor
\State $R \gets \Call{Rank}{R}$ \Comment{rank docs}

\end{algorithmic}
\end{algorithm}

Equation \ref{eq} shows the scoring formulation of \sys{}. Here, $\lambda$ weight is given to the lexical score of the document in consideration and $1-\lambda$ weight is given to the mean lexical scores of the neighbors. In Equation \ref{eq}, ${d_1,d_2,...,d_n}\in N$ is the set of $n$ nearest neighbors to the document in the corpus graph. The final ranking of the documents is established using the newly calculated \sys{} score for each document.  

\begin{equation}
  \sys{}(q,d,D) = \lambda\cdot\text{score}(q,d) + \dfrac{1-\lambda}{n}\cdot\sum_{d_{neigh}\in N}{\text{score}(q,d_{neigh})}\label{eq}
\end{equation}
\\
Figure \ref{fig:sysarch} depicts the system architecture of \sys{}. The corpus graph creation using dense retrieval method is done at indexing time and does not cause any latency overhead at retrieval. For every user query, first BM25 scores are obtained for the complete document corpus. Then as formulated in Equation \ref{eq}, final \sys{} score is assigned to each document as a combination of the document score and mean neighbor score. \sys{} is formally described in Algorithm \ref{alg:fh}. Here, $\Call{Lookup}$ retrieves the precomputed BM25 scores of the neighboring documents. Fusion parameter $\lambda$ and number of neighbors to be considered $n$ are the two key hyper-parameters of the proposed method. The algorithm clearly shows that for each document, its score and neighbor scores are looked up and then final score is calculated using Equation \ref{eq}. 

\section{Experiment}
    
Through extensive experimentation, we address the following research questions:
\begin{description}
\item[RQ1:] Does the neighborhood of a document in a corpus graph provide comprehensive insights regarding relevance of the document to the query?
\item[RQ2:] How does the neighbor score based ranking impact retrieval latency?
\item[RQ3:] How does the effectiveness of the proposed method change with increase in the number of neighbors $n$?
\item[RQ4:] How does the effectiveness of the proposed method change with variation in the fusion parameter $\lambda$?
\item[RQ5:] Does the dataset construction method affect the performance of the proposed method?
\item[RQ6:] Can the optimal fusion parameter be determined through training samples?
\end{description}
 
\noindent We have released the code to reproduce our results for respective research questions here\footnote{https://github.com/Georgetown-IR-Lab/LexBoost}.

\subsection{Datasets and Measures}
To validate the significance of the proposed method through experimentation, we use three publicly available benchmark datasets.
\begin{itemize}
\item \noindent\textbf{TREC 2019 Deep Learning (Passage Subtask).}
%This is the official evaluation query set used in the
This evaluation query set was made available for TREC 2019 Deep Learning shared task \cite{https://doi.org/10.48550/arxiv.2003.07820}. The document corpus is derived from MS MARCO \cite{Bajaj2016Msmarco}. It consists of 43 human-evaluated queries with comprehensive labeling using four relevance grades. This benchmark query set has on an average 215 relevance assessments per query.

\item \noindent\textbf{TREC 2020 Deep Learning (Passage Subtask).}
This evaluation query set was made available for TREC 2020 Deep Learning shared task \cite{https://doi.org/10.48550/arxiv.2102.07662}. The document corpus is derived from MS MARCO \cite{Bajaj2016Msmarco}. It consists of 54 queries with human judgments from NIST annotators. This benchmark query set has on an average 211 relevance assessments per query. Similar to TREC DL 2019, this too has relevance judgements on a four point scale.

\item \noindent\textbf{CORD19/TREC-COVID.} 
Both clinicians and the public has been searching for relevant and reliable information related to COVID-19 since the pandemic. 
%In order to capture the information needs of individuals as well as biomedical researchers, there was a need for relevant and reliable 
TREC COVID is a pandemic retrieval test collection built to create and test retrieval systems for COVID-19 and similar future events
\cite{10.1145/3451964.3451965}. The document set used for this dataset is COVID-19 Open Research Dataset (CORD-19) \cite{wang-etal-2020-cord}. Relevance judgements for the collection of 50 topics are determined by human annotators with biomedical expertise.
\end{itemize}

\subsection{Models and Parameters}

We construct the corpus graph by identifying 16 most similar documents using dense retrieval method TCT-ColBERT-HNP \cite{lin-etal-2021-batch}. Constructing the corpus graph is a one-time process at indexing-stage. The time complexity to build the corpus graph is O($n^2$), while the space complexity is O($n$). 
%Here, we identify 16 most similar documents for each document in the corpus using TCT-ColBERT-HNP. 
We consider 2, 4, 8 and 16 number of closest neighbors ($n$) for each document from the corpus graph for experimentation. This choice is ideal as the impact of significant variation in $n$ on \sys{} performance can be studied. Primarily, the lexical retrieval method used to calculate initial score is BM25 \cite{Robertson2009ThePR}. We define $\lambda$ parameter for fusion between initial BM25 scores and mean neighbor BM25 scores and evaluate for $\lambda$ from 0 to 1 at regular intervals of 0.05. To evaluate the robustness of the \sys{} we also built corpus graph using TAS-B \cite{10.1145/3404835.3462891} and used it to identify neighbors.

We determine and tune the value of fusion parameter $\lambda$ for Convex Combination through validation set given its sample efficiency \cite{10.1145/3596512}. We use TREC DL 19 query set as a validation set for determining optimal value of $\lambda$ parameter. Then, we use this value for \sys{} on TREC DL 20 query set thus validating our optimization approach.
We also evaluate \sys{} with a wide range of $\lambda$ values on all three datasets to understand the impact it has on the nDCG, MAP and Recall evaluations.

\subsection{Baselines and Implementation}

To comprehensively study the performance improvements by \sys{} we compare it with different baseline methods. BM25 is our primary baseline. Further, we also considered PL2, DPH and QLD lexical retrieval methods as baselines to show that \sys{} is agnostic of lexical retrieval method used with it. As corpus graph construction is performed while indexing, it limits the latency overhead during retrieval. Hence, we can directly compare \sys{} Retrieval results with BM25 output in terms of effectiveness. We also show robustness of \sys{} by running experiments across wide range of parameters listed in the above section. Fusion parameter $\lambda$ and number of neighbors considered $n$ are the key hyper-parameters.
Additionally, we also compare \sys{} re-ranking with traditional re-ranking and exhaustive dense retrieval.

% We determine and tune the value of fusion parameter $\lambda$ for Convex Combination through validation set given its sample efficiency \cite{10.1145/3596512}. We use TREC DL 19 query set as a validation set for determining optimal value of $\lambda$ parameter. Then, we use this value for \sys{} on TREC DL 20 query set thus validating our optimization approach.
% We also evaluate \sys{} with a wide range of $\lambda$ values on all three datasets to understand the impact it has on the nDCG, MAP and Recall evaluations.

\begin{figure}
    \centering
\includegraphics[width=\columnwidth]{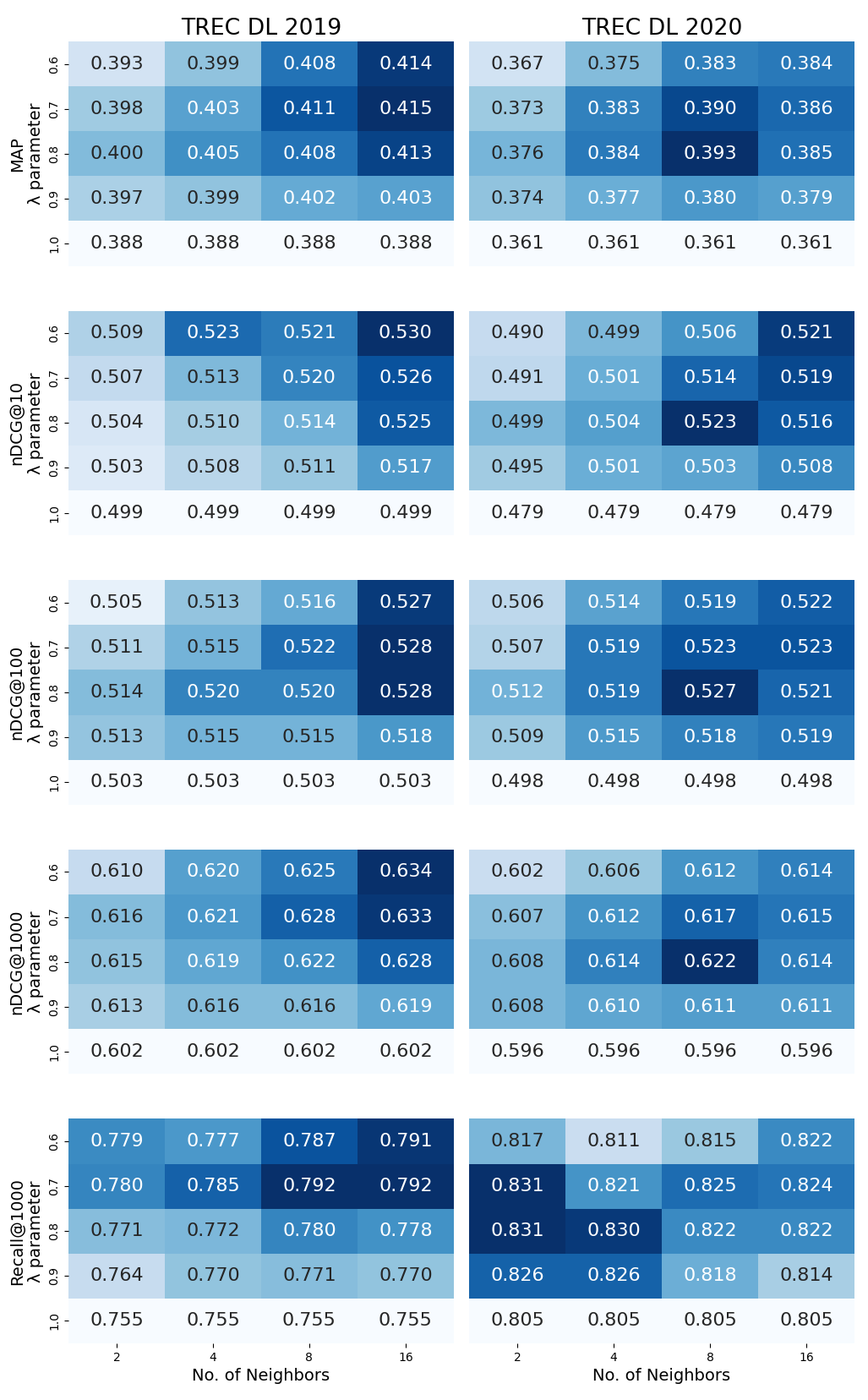}
    \caption{Heat-Maps showing impact of variation in fusion parameter $\lambda$ and no. of neighbors $n$ on \sys{}.}
    \label{fig:heat}
\end{figure}

\begin{table*}
  \caption{\sys{} on TREC DL 2019. $\dagger$ denotes statistically significant improvement (paired t-test: $p < 0.05$). Highest values denoted in bold. In \sys{}(k), k is the number of nearest neighbors.}

  \label{tab:dl19}
  \begin{tabular}{cllccccc}
    \toprule
    $\lambda$ parameter & Method & MAP & nDCG@10 & nDCG@100 & nDCG@1000 & R(rel=2)@1000\\
    \midrule
& BM25 & 0.3877 & 0.4989 & 0.5028 & 0.6023 & 0.7555\\
    \midrule
\multirow{4}{*}{0.7} 
& \sys{}(2) & 0.3977 & 0.5074 & 0.5112 & 0.6158$^\dagger$ & 0.7802$^\dagger$\\
& \sys{}(4) & 0.4032$^\dagger$ & 0.5127 & 0.5151 & 0.621$^\dagger$ & 0.7847$^\dagger$\\
& \sys{}(8) & 0.4107$^\dagger$ & 0.5204 & 0.5224 & 0.6278$^\dagger$ & 0.7918$^\dagger$\\
& \sys{}(16) & \textbf{0.415}$^\dagger$ & 0.526 & 0.5283$^\dagger$ & \textbf{0.6332}$^\dagger$ & \textbf{0.7922}$^\dagger$\\
    \midrule
\multirow{4}{*}{0.8}
& \sys{}(2) & 0.4$^\dagger$ & 0.504 & 0.5142 & 0.6149$^\dagger$ & 0.7707$^\dagger$\\
& \sys{}(4) & 0.4047$^\dagger$ & 0.51 & 0.5204$^\dagger$ & 0.6189$^\dagger$ & 0.7724$^\dagger$\\
& \sys{}(8) & 0.4078$^\dagger$ & 0.5143 & 0.5199$^\dagger$ & 0.6215$^\dagger$ & 0.7796$^\dagger$\\
& \sys{}(16) & 0.413$^\dagger$ & \textbf{0.5251}$^\dagger$ & \textbf{0.5284}$^\dagger$ & 0.6277$^\dagger$ & 0.7784$^\dagger$\\
    \midrule
\multirow{4}{*}{0.9}
& \sys{}(2) & 0.3974$^\dagger$ & 0.503 & 0.5126$^\dagger$ & 0.6131$^\dagger$ & 0.764\\
& \sys{}(4) & 0.3994$^\dagger$ & 0.508 & 0.5146$^\dagger$ & 0.6162$^\dagger$ & 0.7696$^\dagger$\\
& \sys{}(8) & 0.4017$^\dagger$ & 0.5112 & 0.5146$^\dagger$ & 0.6159$^\dagger$ & 0.7705$^\dagger$\\
& \sys{}(16) & 0.4029$^\dagger$ & 0.5171 & 0.5184$^\dagger$ & 0.6188$^\dagger$ & 0.7699$^\dagger$\\
    \midrule
\multirow{4}{*}{0.95}
& \sys{}(2) & 0.3926$^\dagger$ & 0.4985 & 0.5086$^\dagger$ & 0.6088$^\dagger$ & 0.7611\\
& \sys{}(4) & 0.3946$^\dagger$ & 0.5015 & 0.5095$^\dagger$ & 0.6093$^\dagger$ & 0.7606\\
& \sys{}(8) & 0.396$^\dagger$ & 0.5052 & 0.5118$^\dagger$ & 0.6116$^\dagger$ & 0.7634$^\dagger$\\
& \sys{}(16) & 0.3958$^\dagger$ & 0.5091$^\dagger$ & 0.5117$^\dagger$ & 0.6104$^\dagger$ & 0.7616$^\dagger$\\
  \bottomrule
\end{tabular}
\label{tab:dl19}
\end{table*}

\begin{table*}
  \caption{\sys{} on TREC DL 2020. $\dagger$ denotes statistically significant improvement (paired t-test: $p < 0.05$). Highest values denoted in bold. In \sys{}(k), k is the number of nearest neighbors.}
  \label{tab:dl20}
  \begin{tabular}{cllccccc}
    \toprule
    $\lambda$ parameter & Method & MAP & nDCG@10 & nDCG@100 & nDCG@1000 & R(rel=2)@1000\\
    \midrule
& BM25 & 0.3609 & 0.4793 & 0.4984 & 0.5962 & 0.8046\\
\midrule
\multirow{5}{*}{0.7} 
& \sys{}(2) & 0.3726 & 0.4914 & 0.5069 & 0.6068 & 0.8311$^\dagger$\\
& \sys{}(4) & 0.3829$^\dagger$ & 0.5007 & 0.5185$^\dagger$ & 0.6123$^\dagger$ & 0.821$^\dagger$\\
& \sys{}(8) & 0.3897$^\dagger$ & 0.5135$^\dagger$ & 0.5229$^\dagger$ & 0.6175$^\dagger$ & 0.8247$^\dagger$\\
& \sys{}(16) & 0.3862$^\dagger$ & 0.5188$^\dagger$ & 0.523$^\dagger$ & 0.6152$^\dagger$ & 0.8241\\
    \midrule
\multirow{5}{*}{0.8} 
& \sys{}(2) & 0.3757$^\dagger$ & 0.4992 & 0.5116$^\dagger$ & 0.6082$^\dagger$ & \textbf{0.8305}$^\dagger$\\
& \sys{}(4) & 0.3836$^\dagger$ & 0.504$^\dagger$ & 0.5191$^\dagger$ & 0.6143$^\dagger$ & 0.8303$^\dagger$\\
& \sys{}(8) & \textbf{0.3933}$^\dagger$ & \textbf{0.5225}$^\dagger$ & \textbf{0.5272}$^\dagger$ & \textbf{0.6218}$^\dagger$ & 0.8221$^\dagger$\\
& \sys{}(16) & 0.3846$^\dagger$ & 0.5158$^\dagger$ & 0.5214$^\dagger$ & 0.6144$^\dagger$ & 0.8215$^\dagger$\\

    \midrule
\multirow{5}{*}{0.9} 
& \sys{}(2) & 0.3738$^\dagger$ & 0.4946$^\dagger$ & 0.5095$^\dagger$ & 0.6082$^\dagger$ & 0.8263$^\dagger$\\
& \sys{}(4) & 0.377$^\dagger$ & 0.5009$^\dagger$ & 0.5146$^\dagger$ & 0.6104$^\dagger$ & 0.8264$^\dagger$\\
& \sys{}(8) & 0.3797$^\dagger$ & 0.5033$^\dagger$ & 0.5177$^\dagger$ & 0.6114$^\dagger$ & 0.8181$^\dagger$\\
& \sys{}(16) & 0.3795$^\dagger$ & 0.5078$^\dagger$ & 0.5192$^\dagger$ & 0.6114$^\dagger$ & 0.8142$^\dagger$\\
    \midrule
\multirow{5}{*}{0.95} 
& \sys{}(2) & 0.3695$^\dagger$ & 0.4899$^\dagger$ & 0.5075$^\dagger$ & 0.6044$^\dagger$ & 0.8171$^\dagger$\\
& \sys{}(4) & 0.3712$^\dagger$ & 0.4926$^\dagger$ & 0.5081$^\dagger$ & 0.6054$^\dagger$ & 0.8153$^\dagger$\\
& \sys{}(8) & 0.372$^\dagger$ & 0.4941$^\dagger$ & 0.5107$^\dagger$ & 0.6064$^\dagger$ & 0.8126$^\dagger$\\
& \sys{}(16) & 0.3717$^\dagger$ & 0.4957$^\dagger$ & 0.5114$^\dagger$ & 0.6058$^\dagger$ & 0.8084 \\
  \bottomrule
\end{tabular}
\label{tab:dl20}
\end{table*}

\begin{table*}
\caption{\sys{} on TREC COVID. $\dagger$ denotes statistically significant improvement (paired t-test: $p < 0.05$). Highest values denoted in bold. In \sys{}(k), k is the number of nearest neighbors.}
\label{tab:covid}
\begin{tabular}{cllccccc}
\toprule
$\lambda$ parameter & Method & MAP & nDCG@10 & nDCG@100 & nDCG@1000 & R(rel=2)@1000 \\
\midrule
 & BM25 & 0.2525 & 0.6299 & 0.4821 & 0.4191 & 0.4429 \\
\midrule
% \multirow{4}{*}{0.1} 
%  & FastHybrid(2) & 0.2381 & 0.6226 & 0.4744 & 0.4047 & 0.4285 \\
%  & FastHybrid(4) & 0.2466 & 0.6117 & 0.4768 & 0.4148 & 0.4465 \\
%  & FastHybrid(8) & 0.2578 & 0.6208 & 0.4879 & 0.4270 & 0.4620 \\
%  & FastHybrid(16) & 0.2656 & 0.5994 & 0.4785 & 0.4347 & 0.4727$^\dagger$ \\
% \midrule
% \multirow{4}{*}{0.2} 
%  & FastHybrid(2) & 0.2484 & 0.6313 & 0.4823 & 0.4160 & 0.4416 \\
%  & FastHybrid(4) & 0.2606 & 0.6122 & 0.4880 & 0.4288 & 0.4604$^\dagger$ \\
%  & FastHybrid(8) & 0.2737$^\dagger$ & 0.6208 & 0.4991 & 0.4421$^\dagger$ & 0.4773$^\dagger$ \\
%  & FastHybrid(16) & 0.2825$^\dagger$ & 0.6110 & 0.4987 & 0.4536$^\dagger$ & 0.4933$^\dagger$ \\
% \midrule
% \multirow{4}{*}{0.3} 
%  & FastHybrid(2) & 0.2575 & 0.6396 & 0.4895 & 0.4256 & 0.4524 \\
%  & FastHybrid(4) & 0.2711$^\dagger$ & 0.6111 & 0.4966 & 0.4398$^\dagger$ & 0.4744$^\dagger$ \\
%  & FastHybrid(8) & 0.2843$^\dagger$ & 0.6328 & 0.5091$^\dagger$ & 0.4529$^\dagger$ & 0.4870$^\dagger$ \\
%  & FastHybrid(16) & 0.2917$^\dagger$ & 0.6281 & 0.5088$^\dagger$ & 0.4632$^\dagger$ & \textbf{0.5020}$^\dagger$ \\
% \midrule
\multirow{4}{*}{0.4} 
 & \sys{}(2) & 0.2645$^\dagger$ & 0.6472 & 0.4954 & 0.4319$^\dagger$ & 0.4584$^\dagger$ \\
 & \sys{}(4) & 0.2781$^\dagger$ & 0.6257 & 0.5035$^\dagger$ & 0.4471$^\dagger$ & 0.4810$^\dagger$ \\
 & \sys{}(8) & 0.2892$^\dagger$ & 0.6368 & 0.5162$^\dagger$ & 0.4578$^\dagger$ & 0.4915$^\dagger$ \\
 & \sys{}(16) & \textbf{0.2950}$^\dagger$ & 0.6408 & 0.5142$^\dagger$ & \textbf{0.4654}$^\dagger$ & \textbf{0.5017}$^\dagger$ \\
\midrule
\multirow{4}{*}{0.5} 
 & \sys{}(2) & 0.2691$^\dagger$ & 0.6490 & 0.4989$^\dagger$ & 0.4368$^\dagger$ & 0.4637$^\dagger$ \\
 & \sys{}(4) & 0.2813$^\dagger$ & 0.6337 & 0.5075$^\dagger$ & 0.4494$^\dagger$ & 0.4815$^\dagger$ \\
 & \sys{}(8) & 0.2901$^\dagger$ & 0.6402 & 0.5186$^\dagger$ & 0.4588$^\dagger$ & 0.4909$^\dagger$ \\
 & \sys{}(16) & 0.2943$^\dagger$ & 0.6432 & 0.5192$^\dagger$ & 0.4630$^\dagger$ & 0.4963$^\dagger$ \\
\midrule
\multirow{4}{*}{0.6} 
 & \sys{}(2) & 0.2717$^\dagger$ & 0.6528 & 0.5022$^\dagger$ & 0.4402$^\dagger$ & 0.4673$^\dagger$ \\
 & \sys{}(4) & 0.2815$^\dagger$ & 0.6367 & 0.5061$^\dagger$ & 0.4478$^\dagger$ & 0.4776$^\dagger$ \\
 & \sys{}(8) & 0.2881$^\dagger$ & 0.6310 & 0.5177$^\dagger$ & 0.4552$^\dagger$ & 0.4852$^\dagger$ \\
 & \sys{}(16) & 0.2909$^\dagger$ & 0.6483 & \textbf{0.5203}$^\dagger$ & 0.4581$^\dagger$ & 0.4878$^\dagger$ \\
\midrule
\multirow{4}{*}{0.7} 
 & \sys{}(2) & 0.2718$^\dagger$ & 0.6483 & 0.5053$^\dagger$ & 0.4396$^\dagger$ & 0.4659$^\dagger$ \\
 & \sys{}(4) & 0.2790$^\dagger$ & 0.6442 & 0.5083$^\dagger$ & 0.4449$^\dagger$ & 0.4720$^\dagger$ \\
 & \sys{}(8) & 0.2833$^\dagger$ & 0.6444 & 0.5163$^\dagger$ & 0.4496$^\dagger$ & 0.4762$^\dagger$ \\
 & \sys{}(16) & 0.2846$^\dagger$ & 0.6445 & 0.5182$^\dagger$ & 0.4509$^\dagger$ & 0.4788$^\dagger$ \\
\midrule
\multirow{4}{*}{0.8} 
 & \sys{}(2) & 0.2691$^\dagger$ & 0.6496 & 0.5048$^\dagger$ & 0.4365$^\dagger$ & 0.4612$^\dagger$ \\
 & \sys{}(4) & 0.2735$^\dagger$ & 0.6444 & 0.5041$^\dagger$ & 0.4388$^\dagger$ & 0.4639$^\dagger$ \\
 & \sys{}(8) & 0.2754$^\dagger$ & 0.6409 & 0.5093$^\dagger$ & 0.4408$^\dagger$ & 0.4657$^\dagger$ \\
 & \sys{}(16) & 0.2760$^\dagger$ & 0.6465 & 0.5089$^\dagger$ & 0.4424$^\dagger$ & 0.4677$^\dagger$ \\
\midrule
\multirow{4}{*}{0.9} 
 & \sys{}(2) & 0.2632$^\dagger$ & 0.6539 & 0.4978$^\dagger$ & 0.4302$^\dagger$ & 0.4539$^\dagger$ \\
 & \sys{}(4) & 0.2651$^\dagger$ & \textbf{0.6567}$^\dagger$ & 0.5027$^\dagger$ & 0.4312$^\dagger$ & 0.4552$^\dagger$ \\
 & \sys{}(8) & 0.2654$^\dagger$ & 0.6382 & 0.5019$^\dagger$ & 0.4317$^\dagger$ & 0.4566$^\dagger$ \\
 & \sys{}(16) & 0.2654$^\dagger$ & 0.6448$^\dagger$ & 0.4990$^\dagger$ & 0.4320$^\dagger$ & 0.4568$^\dagger$ \\
% \midrule
% \multirow{4}{*}{0.95} 
%  & FastHybrid(2) & 0.2585$^\dagger$ & 0.6433$^\dagger$ & 0.4907$^\dagger$ & 0.4251 & 0.4489$^\dagger$ \\
%  & FastHybrid(4) & 0.2593$^\dagger$ & 0.6437 & 0.4909$^\dagger$ & 0.4253$^\dagger$ & 0.4492$^\dagger$ \\
%  & FastHybrid(8) & 0.2594$^\dagger$ & 0.6420 & 0.4931$^\dagger$ & 0.4251 & 0.4490$^\dagger$ \\
%  & FastHybrid(16) & 0.2592$^\dagger$ & 0.6379 & 0.4902$^\dagger$ & 0.4254$^\dagger$ & 0.4494$^\dagger$ \\
\bottomrule
\end{tabular}
\end{table*}

\section{Results and Analysis}

We now discuss results and their analysis with respect to our research questions. 

\subsection{RQ1 and RQ2: Insights from corpus graph and impact on retrieval}

\begin{figure*}
\centering

nDCG at top 10, 100 and 1000 results

\includegraphics[scale=0.36]{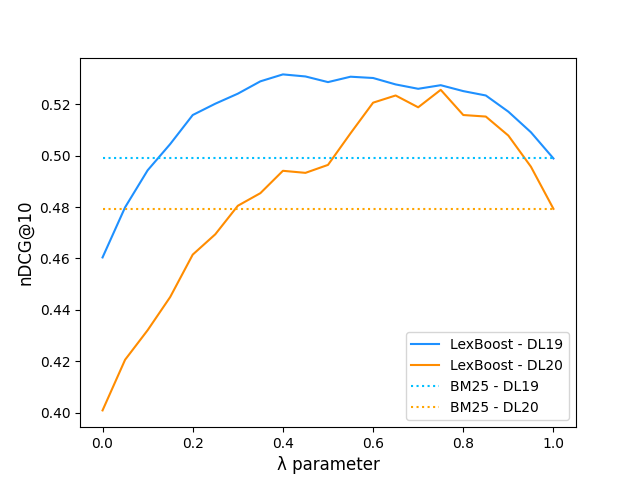}
\includegraphics[scale=0.36]{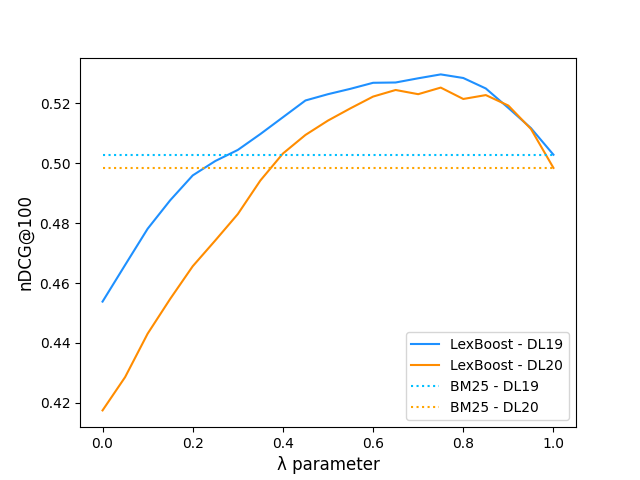}
\includegraphics[scale=0.36]{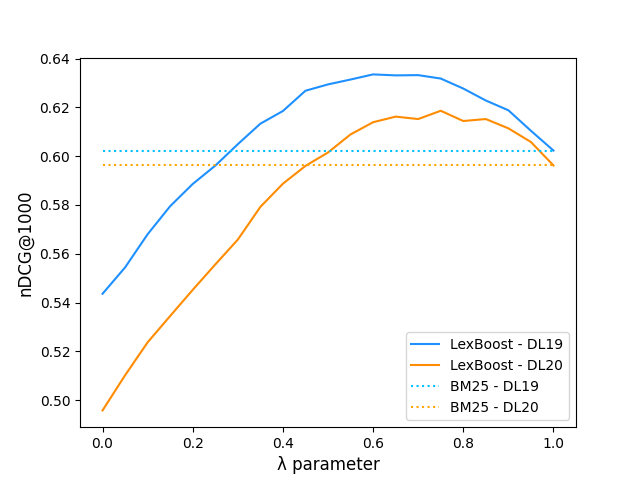}

Mean Average Precision and Recall at first 1000 results

\includegraphics[scale=0.36]{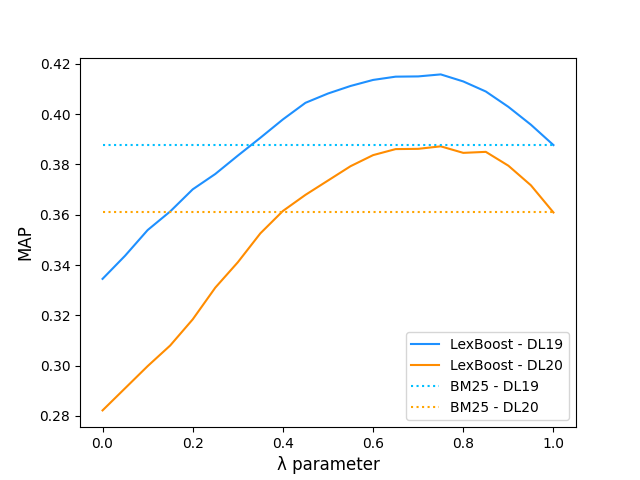}
\includegraphics[scale=0.36]{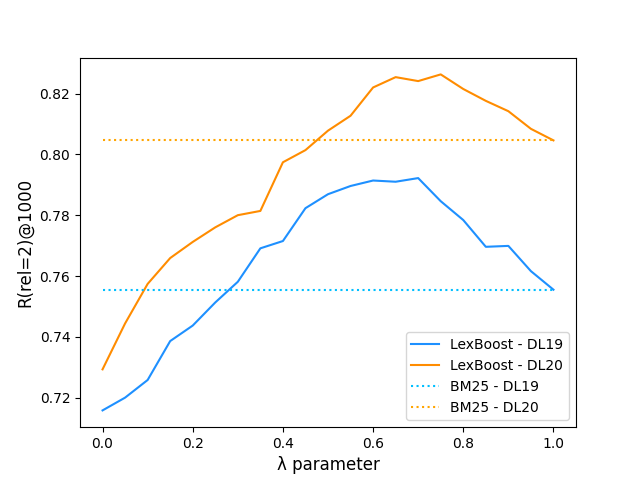}

\caption{Validation based optimization for determination of fusion parameter $\lambda$.}
\label{fig:maingraphs}
\end{figure*}

%RQ1 and RQ2 are about effectively deriving insights from corpus graph and their impact on information retrieval. 
As evident in Table \ref{tab:dl19}, Table \ref{tab:dl20} and Table \ref{tab:covid}, we evaluated results across different datasets, a range of number of nearest neighbors from corpus graph and varying the fusion parameter. We observe statistically significant improvements in retrieval results across all these combinations. Figure \ref{fig:maingraphs} shows these improvements graphically for TREC DL 19 and TREC DL 20 query sets. We evaluated MAP, Recall at 1000 and nDCG at 10, 100 and 1000. 
Using \sys{} on BM25, MAP improved from 0.3877 to 0.415 and from 0.3609 to 0.3933 on TREC DL 19 and 20 respectively. Similarly, using \sys{} on BM25, MAP improved from 0.2525 to 0.2950 on the TREC COVID dataset. \pageenlarge{1}
Further, we also observe that Recall@1000 improved from 0.7555 to 0.7922 and from 0.8046 to 0.8305 on TREC DL 19 and 20 respectively. Similarly, using \sys{} on BM25, Recall@1000 improved from 0.4429 to 0.5020 on the TREC COVID dataset. 
Similarly, statistically significant improvements were observed on nDCG@100 and nDCG@1000 with \sys{} across the three datasets under consideration. 

\begin{figure}
\centering

\includegraphics[width=0.8\columnwidth]{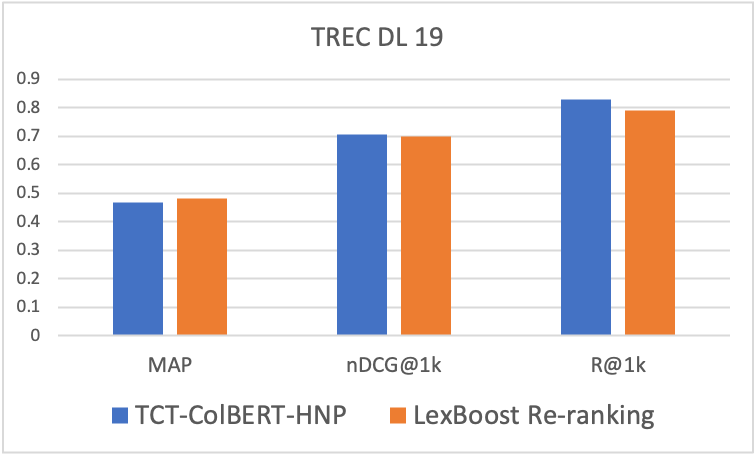}

\includegraphics[width=0.8\columnwidth]{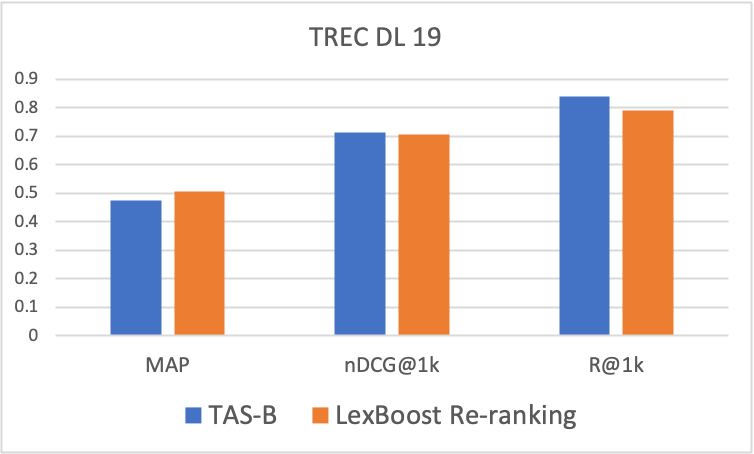}
%scale=0.36

\caption{Comparison of \sys{} Re-ranking with exhaustive dense retrieval with TCT-Colbert-HNP and TAS-B on TREC DL 2019 query set.}
\label{fig:den}
\end{figure}

%\pageenlarge{1}
To test the applicability of \sys{}, we evaluated it over four popular and effective lexical retrieval methods. As evident in Table \ref{tab:base}, \sys{} shows statistically significant improvements over BM25, PL2, DPH and QLD baselines. Hence, we infer that \sys{} is robust and can be applied over a wide range of lexical retrieval methods. \pageenlarge{1} 
Additionally, we created the corpus graph using multiple dense retrieval methods namely: TCT-ColBERT-HNP and TAS-B. \sys{} shows statistically significant improvements using both corpus graphs when evaluated on TREC DL 19 and 20 as evident in Table \ref{tab:tasb}. This shows that \sys{} is robust and agnostic to the dense method used to build the corpus graph. 
Further, we evaluate re-ranking using \sys{}. Here, we use a dense retrieval method on top of \sys{} to re-rank top 1000 documents. As evident in Table \ref{tab:rr}, we observe statistically significant improvements over traditional re-ranking pipelines. This experiment was conducted on TREC DL 19 and 20 using TCT-ColBERT-HNP and TAS-B dense retrieval methods. Further, as evident in Figure \ref{fig:den} we compare \sys{} re-ranking with exhaustive dense retrieval. \pageenlarge{2}
We do not consider exhaustive dense retrieval to be a baseline as it not equivalent to \sys{} re-ranking from latency point of view. In Figure \ref{fig:den}, we notice that \sys{} re-ranking outperforms dense retrieval in MAP and show comparable performance in other metrics. 
This result is important as \sys{} re-ranking is significantly more efficient than exhaustive dense retrieval.
   
Most importantly, these improvements come with negligible additional latency overheads. \pageenlarge{1}
The corpus graph is built and neighbors are identified while indexing the documents. Hence, \sys{} enables to use dense retrieval based document similarity insights effectively during retrieval limiting latency overheads
%Further, it effectively derives and uses the insights from corpus graph neighbors 
addressing RQ1 and RQ2.

\begin{table}
\caption{\sys{} on MS MARCO TREC DL 20. $\dagger$ denotes statistically significant improvement (paired t-test: $p < 0.05$). Highest values denoted in bold. In \sys{}(k), k is the number of nearest neighbors. Top fusion parameter $\lambda$ values determined using TREC DL 19 query set.\\}
\label{tab:base}
\begin{tabular}{cllccc}
\toprule
Method & $\lambda$ & MAP & nDCG@1k & R(rel=2)@1k \\
\midrule
BM25 & & 0.3609 & 0.5962 & 0.8046 \\
\midrule
\multirow{3}{*}{\sys{}(16)} 
 & 0.7 & \textbf{0.3862}$^\dagger$ & \textbf{0.6152}$^\dagger$ & \textbf{0.8241} \\
 & 0.8 & 0.3846$^\dagger$ & 0.6144$^\dagger$ & 0.8215$^\dagger$ \\
 & 0.9 & 0.3795$^\dagger$ & 0.6114$^\dagger$ & 0.8142$^\dagger$ \\
\midrule
PL2 & & 0.3227 & 0.5609 & 0.7772 \\
\midrule
\multirow{3}{*}{\sys{}(16)} 
 & 0.7 & \textbf{0.3508}$^\dagger$ & \textbf{0.5874}$^\dagger$ & \textbf{0.8011}$^\dagger$ \\
 & 0.8 & 0.3464$^\dagger$ & 0.5823$^\dagger$ & 0.7940 \\
 & 0.9 & 0.3363$^\dagger$ & 0.5747$^\dagger$ & 0.7939$^\dagger$ \\
\midrule
DPH & & 0.3363 & 0.5704 & 0.7980 \\
\midrule
\multirow{3}{*}{\sys{}(16)} 
 & 0.7 & \textbf{0.3645}$^\dagger$ & \textbf{0.5970}$^\dagger$ & \textbf{0.8195}$^\dagger$ \\
 & 0.8 & 0.3638$^\dagger$ & 0.5951$^\dagger$ & 0.8123$^\dagger$ \\
 & 0.9 & 0.3532$^\dagger$ & 0.5855$^\dagger$ & 0.8052$^\dagger$ \\
\midrule
QLD & & 0.3580 & 0.5870 & 0.8125 \\
\midrule
\multirow{3}{*}{\sys{}(16)} 
 & 0.7 & \textbf{0.3987}$^\dagger$ & \textbf{0.6198}$^\dagger$ & \textbf{0.8347}$^\dagger$ \\
 & 0.8 & 0.3933$^\dagger$ & 0.6165$^\dagger$ & 0.8325$^\dagger$ \\
 & 0.9 & 0.3790$^\dagger$ & 0.6045$^\dagger$ & 0.8258$^\dagger$ \\
\bottomrule
\end{tabular}
\end{table}
\begin{table}
\caption{\sys{} on MS MARCO TREC DL 19 and 20. $\dagger$ denotes statistically significant improvement (paired t-test: $p < 0.05$). Highest values denoted in bold. In \sys{}(k)(cg), k is the number of nearest neighbors and cg is the method used to construct corpus graph. Corpus graph constructed with HNP: TCT-ColBERT-HNP and TAS: TAS-B.}
\label{tab:tasb}
\begin{tabular}{cllcc}
\toprule
Method & $\lambda$ & MAP & nDCG@1k & R(rel=2)@1k \\
\midrule
&&DL 19&&\\
\midrule
BM25 & & 0.3877 & 0.6023 & 0.7555 \\
\midrule
& 0.7 & \textbf{0.4150}$^\dagger$ & \textbf{0.6332}$^\dagger$ & \textbf{0.7922}$^\dagger$ \\
\sys{}(16)(HNP) & 0.8 & 0.4130$^\dagger$ & 0.6277$^\dagger$ & 0.7784$^\dagger$ \\
& 0.9 & 0.4029$^\dagger$ & 0.6188$^\dagger$ & 0.7699$^\dagger$ \\
\midrule
& 0.7 & \textbf{0.4147}$^\dagger$ & \textbf{0.6355}$^\dagger$ & \textbf{0.7896}$^\dagger$ \\
\sys{}(16)(TAS) & 0.8 & 0.4123$^\dagger$ & 0.6272$^\dagger$ & 0.7766$^\dagger$ \\
& 0.9 & 0.4025$^\dagger$ & 0.6196$^\dagger$ & 0.7735$^\dagger$ \\
\midrule
&&DL 20&&\\
\midrule
BM25 & & 0.3609 & 0.5962 & 0.8046 \\
\midrule
& 0.7 & \textbf{0.3862}$^\dagger$ & \textbf{0.6152}$^\dagger$ & \textbf{0.8241} \\
\sys{}(16)(HNP) & 0.8 & 0.3846$^\dagger$ & 0.6144$^\dagger$ & 0.8215$^\dagger$ \\
& 0.9 & 0.3795$^\dagger$ & 0.6114$^\dagger$ & 0.8142$^\dagger$ \\
\midrule
& 0.7 & \textbf{0.3966}$^\dagger$ & \textbf{0.6246}$^\dagger$ & \textbf{0.8284}$^\dagger$ \\
\sys{}(16)(TAS) & 0.8 & 0.3892$^\dagger$ & 0.6190$^\dagger$ & 0.8267$^\dagger$ \\
& 0.9 & 0.3791$^\dagger$ & 0.6045$^\dagger$ & 0.8152$^\dagger$ \\
\bottomrule
\end{tabular}
\end{table}
\begin{table}
\caption{Re-ranking using \sys{} on MS MARCO TREC DL 19 and 20. $\dagger$ denotes statistically significant improvement (paired t-test: $p < 0.05$). Highest values denoted in bold. In \sys{}(k)(m), k is the number of nearest neighbors and m is the dense retrieval method used for re-ranking. Corpus graph constructed with TCT-ColBERT-HNP. For m - HNP: TCT-ColBERT-HNP and TAS: TAS-B.\\}
\label{tab:rr}
\begin{tabular}{cllcc}
\toprule
Method & $\lambda$ & MAP & nDCG@1k & R(rel=2)@1k \\
\midrule
&&DL 19&&\\
\midrule
BM25 >> HNP & & 0.4643 & 0.6786 & 0.7555 \\
\midrule
& 0.7 & \textbf{0.4832}$^\dagger$ & \textbf{0.7005}$^\dagger$ & \textbf{0.7922}$^\dagger$ \\
\sys{}(16)(HNP) & 0.8 & 0.4768$^\dagger$ & 0.6933$^\dagger$ & 0.7784$^\dagger$ \\
& 0.9 & 0.4728$^\dagger$ & 0.6887$^\dagger$ & 0.7699$^\dagger$ \\
\midrule
BM25 >> TAS & & 0.4888 & 0.6842 & 0.7555 \\
\midrule
& 0.7 & \textbf{0.5070}$^\dagger$ & \textbf{0.7057}$^\dagger$ & \textbf{0.7922}$^\dagger$ \\
\sys{}(16)(TAS) & 0.8 & 0.4995 & 0.6982$^\dagger$ & 0.7784$^\dagger$ \\
& 0.9 & 0.4971$^\dagger$ & 0.6946$^\dagger$ & 0.7699$^\dagger$ \\
\midrule
&&DL 20&&\\
\midrule
BM25 >> HNP & & 0.4696 & 0.6854 & 0.8048 \\
\midrule
& 0.7 & \textbf{0.4779} & \textbf{0.6967} & \textbf{0.8241} \\
\sys{}(16)(HNP) & 0.8 & 0.4763 & 0.6953$^\dagger$ & 0.8215$^\dagger$ \\
& 0.9 & 0.4731$^\dagger$ & 0.6911$^\dagger$ & 0.8142$^\dagger$ \\
\midrule
BM25 >> TAS & & 0.4878 & 0.6912 & 0.8048 \\
\midrule
& 0.7 & \textbf{0.4939} & \textbf{0.7023} & \textbf{0.8241} \\
\sys{}(16)(TAS) & 0.8 & 0.4933 & 0.7011$^\dagger$ & 0.8215$^\dagger$ \\
& 0.9 & 0.4903$^\dagger$ & 0.6968$^\dagger$ & 0.8142$^\dagger$ \\
\bottomrule
\end{tabular}
\end{table}

\subsection{RQ3: Robustness across number of neighbors considered}

%RQ3 is about working of proposed approach irrespective of selection number of neighbors to derive insights. 
As evident in Table \ref{tab:dl19}, Table \ref{tab:dl20} and Table \ref{tab:covid}, we evaluated \sys{} considering a range of nearest neighbors (2, 4, 8, 16) of the target document from the corpus graph. Neighbors are the documents in the corpus graph that are most similar to the target document. We observe statistically significant improvements irrespective of number of neighbors selected - with a general trend of higher improvements and better statistical significance with more neighbors in consideration. Thus, the increasing number of neighbors under consideration deliver better insights. 
Trends can be better understood with the heat-maps shown in Figure \ref{fig:heat}. The left set of heat-maps is for TREC DL 19 query set while the right set of heat-maps is for TREC DL 20 query set. The heat-maps depict combinations of fusion parameter $\lambda$ and number of neighbors considered $n$. For each of the query set we have five heat-maps for the five metrics we are evaluating. Darker the shade of blue, better is the performance of \sys{} for that specific combination in that metric. As evident in Figure \ref{fig:heat}, the general trend is - with increase in the number of neighbors considered the performance improves. The bottom-most row in each heat-map is for fusion parameter $\lambda = 1$ which is equivalent to the baseline BM25. The significant improvements in performance across the variety of number of neighbors considered is also clearly evident by the drastic change in color between the bottom-most rows and others.
This establishes robustness across the number of nearest neighbors considered from the corpus graph for \sys{} - hence addressing RQ3.

\subsection{RQ4: Robustness across variation in fusion parameter $\lambda$}

The fusion parameter $\lambda$ decides the role played by neighbors in relevance of the target document to the user query. We evaluated \sys{} for fusion parameter $\lambda$ values from 0 to 1 at regular intervals of 0.05 as can be seen in Figure \ref{fig:maingraphs}. Here, the five plots are for five metrics under consideration. Part of \sys{} curve above respective baseline gives the range of $\lambda$ parameter values leading to improvement in performance.  
Further, some of the $\lambda$ values leading to highest improvements are evident in Table \ref{tab:dl19}, Table \ref{tab:dl20} and Table \ref{tab:covid}. 
We evaluated for these values across three datasets and varying number of nearest neighbors. We observed statistically significant improvements across a wide array of these combinations. The trends are evident in more detail in the heat-maps shown in Figure \ref{fig:heat}. On the y-axis for each heat-map the fusion parameter $\lambda$ varies from 0.6 to 1. Darker the shade of blue - higher is the performance of \sys{} for that combination. It is evident that for a wide range of $\lambda$ values significant improvements are observed. Further, the two different query sets are in sync with respect to the optimal $\lambda$ value as evident in Figure \ref{fig:heat}. This shows effectiveness of our proposed method \sys{} for a range of fusion parameter values establishing robustness and hence addressing RQ4. 

\subsection{RQ5: Robustness across different datasets under consideration}
RQ5 is about \sys{} working effectively across different datasets and independent of dataset preparation methods. As evident in Table \ref{tab:dl19}, Table \ref{tab:dl20} and Table \ref{tab:covid}, we evaluated \sys{} across multiple datasets with different construction mechanisms. 
MS MARCO v1 passage ranking dataset was constructed by taking union of top passage lists for a large set of queries \cite{soboroff2021overview}. On the other hand, TREC-COVID dataset uses documents from CORD-19 which is a large set of scholarly articles about COVID \cite{10.1145/3451964.3451965}. 
In each case we found statistically significant improvements for a wide range of combinations of fusion parameter $\lambda$ and number of neighbors considered. This shows robustness of \sys{} across different datasets, hence addressing RQ5.

\subsection{RQ6: Tuning and optimization for optimal fusion parameter $\lambda$}
  
%RQ6 is about optimization. Fusion parameter is the key of optimization. 
Figure \ref{fig:maingraphs} shows performance of \sys{} in five metrics, namely nDCG@10, nDCG@100, nDCG@1000, MAP and Recall(rel=2)@1000 for varying fusion parameter $\lambda$ values. This evaluation is done using the TREC DL 19 and TREC DL 20 query sets. The points for the curve have been plotted for value of $\lambda$ from 0 to 1 at regular intervals of 0.05. The BM25 baseline is shown by a flat line of the same color. In Figure \ref{fig:maingraphs}, it is evident that for a wide range of $\lambda$ values \sys{} leads to a better performance than the BM25 baseline. Further, we also note that both the TREC DL 19 and TREC DL 20 curves are in sync for all the five metrics in the respective graphs. This validates the effective use of training set in determining optimal fusion parameter $\lambda$ value - addressing RQ6.

\section{Conclusions and Future Directions}
We proposed \sys{} method which utilizes location of the target document in the corpus graph to gain valuable semantic insights from the neighboring documents along with their BM25 scores to determine the final score to be assigned. The enrichment resulted through semantic insights contributes to the increase in effectiveness. \sys{} provides a mechanism of effectively utilizing dense retrieval based similarity derived from the corpus graph with virtually no additional latency overheads at query time.
This shows statistically significant improvements in precision and recall. The method is robust across number of neighbors considered, variation in fusion parameter $\lambda$ and multiple datasets. Overall robustness and improvements with virtually no additional cost makes \sys{} very impactful. \sys{} re-ranking also shows significant improvements over traditional re-ranking results. Further, \sys{} re-ranking shows comparable performance to high-latency exhaustive dense retrieval.

As a future work, \sys{} could be further extended for Cross-Lingual Information Retrieval (CLIR) and Multi-Lingual Information Retrieval (MLIR) settings. Additionally, the proposed approach can be extended to scenario which has user history. Here, a joint document-query graph can be built for stronger insights.
We also plan to evaluate \sys{} architecture on more efficiently built graphs which use approximation methods for similarity calculations. We would like leave dynamic (runtime) tuning of fusion parameter $\lambda$ for the future work. 

%possibilities of deriving dynamic insights through dynamic mapping can also be explored. Even association among multiple corpus graphs to derive hidden insights for Multi-language scenarios could lead to an interesting research directions opening new possibilities of future applications of \sys{} Retrieval. Additionally, we would also like to leave the experimentation with construction of corpus graph using different approaches for future work.

% \begin{table*}
%   \caption{Some Typical Commands}
%   \label{tab:commands}
%   \begin{tabular}{ccl}
%     \toprule
%     Command &A Number & Comments\\
%     \midrule
%     \texttt{{\char'134}author} & 100& Author \\
%     \texttt{{\char'134}table}& 300 & For tables\\
%     \texttt{{\char'134}table*}& 400& For wider tables\\
%     \bottomrule
%   \end{tabular}
% \end{table*}

\bibliographystyle{ACM-Reference-Format}
%\bibliography{sample-base}
\bibliography{sample-new}


\begin{thebibliography}{47}

%%% the Web address.

\ifx \showCODEN    \undefined \def \showCODEN     #1{\unskip}     \fi
\ifx \showDOI      \undefined \def \showDOI       #1{#1}\fi
\ifx \showISBNx    \undefined \def \showISBNx     #1{\unskip}     \fi
\ifx \showISBNxiii \undefined \def \showISBNxiii  #1{\unskip}     \fi
\ifx \showISSN     \undefined \def \showISSN      #1{\unskip}     \fi
\ifx \showLCCN     \undefined \def \showLCCN      #1{\unskip}     \fi
\ifx \shownote     \undefined \def \shownote      #1{#1}          \fi
\ifx \showarticletitle \undefined \def \showarticletitle #1{#1}   \fi
\ifx \showURL      \undefined \def \showURL       {\relax}        \fi
% invisible to TeX
\providecommand\bibfield[2]{#2}
\providecommand\bibinfo[2]{#2}
\providecommand\natexlab[1]{#1}
\providecommand\showeprint[2][]{arXiv:#2}

\bibitem[Abdul-Jaleel et~al\mbox{.}(2004)]%
        {abduljaleel2004umass}
\bibfield{author}{\bibinfo{person}{Nasreen Abdul-Jaleel} {et~al\mbox{.}}} \bibinfo{year}{2004}\natexlab{}.
\newblock \showarticletitle{UMass at {TREC} 2004: Novelty and {HARD}}.
\newblock \bibinfo{journal}{\emph{Computer Science Department Faculty Publication Series}}  \bibinfo{volume}{189.} (\bibinfo{year}{2004}).
\newblock


\bibitem[Acquavia et~al\mbox{.}(2023)]%
        {10.1145/3573128.3604896}
\bibfield{author}{\bibinfo{person}{Antonio Acquavia} {et~al\mbox{.}}} \bibinfo{year}{2023}\natexlab{}.
\newblock \showarticletitle{Static Pruning for Multi-Representation Dense Retrieval}. In \bibinfo{booktitle}{\emph{Proceedings of the ACM Symposium on Document Engineering 2023}} (Limerick, Ireland) \emph{(\bibinfo{series}{DocEng '23})}.
\newblock
\showISBNx{9798400700279}


\bibitem[Amati and Van~Rijsbergen(2002)]%
        {10.1145/582415.582416}
\bibfield{author}{\bibinfo{person}{Gianni Amati} {and} \bibinfo{person}{Cornelis~Joost Van~Rijsbergen}.} \bibinfo{year}{2002}\natexlab{}.
\newblock \showarticletitle{Probabilistic models of information retrieval based on measuring the divergence from randomness}.
\newblock \bibinfo{journal}{\emph{ACM Trans. Inf. Syst.}} \bibinfo{volume}{20}, \bibinfo{number}{4} (\bibinfo{date}{oct} \bibinfo{year}{2002}), \bibinfo{pages}{357–389}.
\newblock
\showISSN{1046-8188}


\bibitem[Bajaj et~al\mbox{.}(2016)]%
        {Bajaj2016Msmarco}
\bibfield{author}{\bibinfo{person}{Payal Bajaj} {et~al\mbox{.}}} \bibinfo{year}{2016}\natexlab{}.
\newblock \showarticletitle{MS MARCO: A Human Generated MAchine Reading COmprehension Dataset}. In \bibinfo{booktitle}{\emph{InCoCo@NIPS}}.
\newblock


\bibitem[Bartell et~al\mbox{.}(1998)]%
        {qld}
\bibfield{author}{\bibinfo{person}{Brian~T. Bartell} {et~al\mbox{.}}} \bibinfo{year}{1998}\natexlab{}.
\newblock \showarticletitle{{Optimizing similarity using multi‐query relevance feedback}}.
\newblock \bibinfo{journal}{\emph{Journal of the American Society for Information Science}} \bibinfo{volume}{49}, \bibinfo{number}{8} (\bibinfo{year}{1998}).
\newblock


\bibitem[Brown et~al\mbox{.}(2020)]%
        {NEURIPS2020_1457c0d6}
\bibfield{author}{\bibinfo{person}{Tom Brown} {et~al\mbox{.}}} \bibinfo{year}{2020}\natexlab{}.
\newblock \showarticletitle{Language Models are Few-Shot Learners}. In \bibinfo{booktitle}{\emph{Advances in Neural Information Processing Systems}}, Vol.~\bibinfo{volume}{33}. \bibinfo{publisher}{Curran Associates, Inc.}, \bibinfo{pages}{1877--1901}.
\newblock


\bibitem[Bruch et~al\mbox{.}(2023)]%
        {10.1145/3596512}
\bibfield{author}{\bibinfo{person}{Sebastian Bruch} {et~al\mbox{.}}} \bibinfo{year}{2023}\natexlab{}.
\newblock \showarticletitle{An Analysis of Fusion Functions for Hybrid Retrieval}.
\newblock \bibinfo{journal}{\emph{ACM Trans. Inf. Syst.}} \bibinfo{volume}{42}, \bibinfo{number}{1}, Article \bibinfo{articleno}{20} (\bibinfo{date}{aug} \bibinfo{year}{2023}).
\newblock
\showISSN{1046-8188}


\bibitem[Carpineto and Romano(2012)]%
        {10.1145/2071389.2071390}
\bibfield{author}{\bibinfo{person}{Claudio Carpineto} {and} \bibinfo{person}{Giovanni Romano}.} \bibinfo{year}{2012}\natexlab{}.
\newblock \showarticletitle{A Survey of Automatic Query Expansion in Information Retrieval}.
\newblock \bibinfo{journal}{\emph{ACM Comput. Surv.}} \bibinfo{volume}{44}, \bibinfo{number}{1}, Article \bibinfo{articleno}{1} (\bibinfo{date}{jan} \bibinfo{year}{2012}).
\newblock
\showISSN{0360-0300}


\bibitem[Chatterjee et~al\mbox{.}(2024)]%
        {10.1007/978-3-031-56027-9_13}
\bibfield{author}{\bibinfo{person}{Shubham Chatterjee} {et~al\mbox{.}}} \bibinfo{year}{2024}\natexlab{}.
\newblock \showarticletitle{DREQ: Document Re-ranking Using Entity-Based Query Understanding}. In \bibinfo{booktitle}{\emph{Advances in Information Retrieval: 46th European Conference on Information Retrieval, ECIR 2024, Glasgow, UK, March 24–28, 2024.}}
\newblock
\showISBNx{978-3-031-56027-9}


\bibitem[Cohen et~al\mbox{.}(2024)]%
        {10.1007/978-3-031-56063-7_16}
\bibfield{author}{\bibinfo{person}{Nachshon Cohen} {et~al\mbox{.}}} \bibinfo{year}{2024}\natexlab{}.
\newblock \showarticletitle{InDi: Informative and Diverse Sampling for Dense Retrieval}. In \bibinfo{booktitle}{\emph{Advances in Information Retrieval: 46th European Conference on Information Retrieval, ECIR 2024, Glasgow, UK, March 24–28, 2024.}}
\newblock
\showISBNx{978-3-031-56063-7}


\bibitem[Craswell et~al\mbox{.}(2020)]%
        {https://doi.org/10.48550/arxiv.2003.07820}
\bibfield{author}{\bibinfo{person}{Nick Craswell} {et~al\mbox{.}}} \bibinfo{year}{2020}\natexlab{}.
\newblock \bibinfo{title}{Overview of the TREC 2019 deep learning track}.
\newblock
\newblock
\urldef\tempurl%
\url{https://doi.org/10.48550/ARXIV.2003.07820}
\showDOI{\tempurl}


\bibitem[Craswell et~al\mbox{.}(2021)]%
        {https://doi.org/10.48550/arxiv.2102.07662}
\bibfield{author}{\bibinfo{person}{Nick Craswell} {et~al\mbox{.}}} \bibinfo{year}{2021}\natexlab{}.
\newblock \bibinfo{title}{Overview of the TREC 2020 deep learning track}.
\newblock
\newblock
\urldef\tempurl%
\url{https://doi.org/10.48550/ARXIV.2102.07662}
\showDOI{\tempurl}


\bibitem[Devlin et~al\mbox{.}(2019)]%
        {devlin-etal-2019-bert}
\bibfield{author}{\bibinfo{person}{Jacob Devlin} {et~al\mbox{.}}} \bibinfo{year}{2019}\natexlab{}.
\newblock \showarticletitle{{BERT}: Pre-training of Deep Bidirectional Transformers for Language Understanding}. In \bibinfo{booktitle}{\emph{Proceedings of the 2019 Conference of the North {A}merican Chapter of the Association for Computational Linguistics)}}. \bibinfo{address}{Minneapolis, Minnesota}, \bibinfo{pages}{4171--4186}.
\newblock


\bibitem[Goker and Davies(2009)]%
        {irbook}
\bibfield{author}{\bibinfo{person}{Ayse Goker} {and} \bibinfo{person}{John Davies}.} \bibinfo{year}{2009}\natexlab{}.
\newblock \showarticletitle{Information Retrieval: Searching in the 21st Century}.
\newblock  (\bibinfo{date}{10} \bibinfo{year}{2009}).
\newblock
\showISBNx{9780470027622}
\urldef\tempurl%
\url{https://doi.org/10.1002/9780470033647}
\showDOI{\tempurl}


\bibitem[Hofst\"{a}tter et~al\mbox{.}(2021)]%
        {10.1145/3404835.3462891}
\bibfield{author}{\bibinfo{person}{Sebastian Hofst\"{a}tter} {et~al\mbox{.}}} \bibinfo{year}{2021}\natexlab{}.
\newblock \showarticletitle{Efficiently Teaching an Effective Dense Retriever with Balanced Topic Aware Sampling}. In \bibinfo{booktitle}{\emph{Proceedings of the 44th International ACM SIGIR Conference on Research and Development in Information Retrieval}} (Virtual Event, Canada) \emph{(\bibinfo{series}{SIGIR '21})}. \bibinfo{pages}{113–122}.
\newblock
\showISBNx{9781450380379}


\bibitem[Huang et~al\mbox{.}(2013)]%
        {10.1145/2505515.2505665}
\bibfield{author}{\bibinfo{person}{Po-Sen Huang} {et~al\mbox{.}}} \bibinfo{year}{2013}\natexlab{}.
\newblock \showarticletitle{Learning deep structured semantic models for web search using clickthrough data}. In \bibinfo{booktitle}{\emph{Proceedings of the 22nd ACM International Conference on Information \& Knowledge Management}} (San Francisco, California, USA) \emph{(\bibinfo{series}{CIKM '13})}. \bibinfo{pages}{2333–2338}.
\newblock
\showISBNx{9781450322638}


\bibitem[Jardine and van Rijsbergen(1971)]%
        {Jardine1971TheUO}
\bibfield{author}{\bibinfo{person}{N. Jardine} {and} \bibinfo{person}{C.~J. van Rijsbergen}.} \bibinfo{year}{1971}\natexlab{}.
\newblock \showarticletitle{The use of hierarchic clustering in information retrieval}.
\newblock \bibinfo{journal}{\emph{Inf. Storage Retr.}}  \bibinfo{volume}{7} (\bibinfo{year}{1971}), \bibinfo{pages}{217--240}.
\newblock


\bibitem[Kulkarni et~al\mbox{.}(2023a)]%
        {10.1145/3573128.3609340}
\bibfield{author}{\bibinfo{person}{Hrishikesh Kulkarni} {et~al\mbox{.}}} \bibinfo{year}{2023}\natexlab{a}.
\newblock \showarticletitle{Genetic Generative Information Retrieval}. In \bibinfo{booktitle}{\emph{Proceedings of the ACM Symposium on Document Engineering 2023}} (Limerick, Ireland) \emph{(\bibinfo{series}{DocEng '23})}. Article \bibinfo{articleno}{8}, \bibinfo{numpages}{4}~pages.
\newblock
\showISBNx{9798400700279}


\bibitem[Kulkarni et~al\mbox{.}(2023b)]%
        {10.1145/3539618.3591715}
\bibfield{author}{\bibinfo{person}{Hrishikesh Kulkarni} {et~al\mbox{.}}} \bibinfo{year}{2023}\natexlab{b}.
\newblock \showarticletitle{Lexically-Accelerated Dense Retrieval}. In \bibinfo{booktitle}{\emph{Proceedings of the 46th International ACM SIGIR Conference on Research and Development in Information Retrieval}} (Taipei, Taiwan) \emph{(\bibinfo{series}{SIGIR '23})}. \bibinfo{pages}{152–162}.
\newblock
\showISBNx{9781450394086}


\bibitem[Li et~al\mbox{.}(2023)]%
        {10.1145/3539618.3591651}
\bibfield{author}{\bibinfo{person}{Haitao Li} {et~al\mbox{.}}} \bibinfo{year}{2023}\natexlab{}.
\newblock \showarticletitle{Constructing Tree-based Index for Efficient and Effective Dense Retrieval}. In \bibinfo{booktitle}{\emph{Proceedings of the 46th International ACM SIGIR Conference on Research and Development in Information Retrieval}} (Taipei, Taiwan) \emph{(\bibinfo{series}{SIGIR '23})}. \bibinfo{pages}{131–140}.
\newblock
\showISBNx{9781450394086}


\bibitem[Lin et~al\mbox{.}(2021)]%
        {lin-etal-2021-batch}
\bibfield{author}{\bibinfo{person}{Sheng-Chieh Lin} {et~al\mbox{.}}} \bibinfo{year}{2021}\natexlab{}.
\newblock \showarticletitle{In-Batch Negatives for Knowledge Distillation with Tightly-Coupled Teachers for Dense Retrieval}. In \bibinfo{booktitle}{\emph{Proceedings of the 6th Workshop on Representation Learning for NLP (RepL4NLP-2021)}}. \bibinfo{publisher}{Association for Computational Linguistics}, \bibinfo{address}{Online}, \bibinfo{pages}{163--173}.
\newblock


\bibitem[Liu(2007)]%
        {liu2007web}
\bibfield{author}{\bibinfo{person}{Bing Liu}.} \bibinfo{year}{2007}\natexlab{}.
\newblock \showarticletitle{Web usage mining}.
\newblock \bibinfo{journal}{\emph{Web Data Mining: Exploring Hyperlinks, Contents, and Usage Data}} (\bibinfo{year}{2007}), \bibinfo{pages}{449--483}.
\newblock


\bibitem[MacAvaney et~al\mbox{.}(2022)]%
        {macavaney:cikm2022-adaptive}
\bibfield{author}{\bibinfo{person}{Sean MacAvaney} {et~al\mbox{.}}} \bibinfo{year}{2022}\natexlab{}.
\newblock \showarticletitle{Adaptive Re-Ranking with a Corpus Graph}. In \bibinfo{booktitle}{\emph{31st ACM International Conference on Information and Knowledge Management}}.
\newblock


\bibitem[Malkov and Yashunin(2020)]%
        {8594636}
\bibfield{author}{\bibinfo{person}{Yu~A. Malkov} {and} \bibinfo{person}{D.~A. Yashunin}.} \bibinfo{year}{2020}\natexlab{}.
\newblock \showarticletitle{Efficient and Robust Approximate Nearest Neighbor Search Using Hierarchical Navigable Small World Graphs}.
\newblock \bibinfo{journal}{\emph{IEEE Transactions on Pattern Analysis and Machine Intelligence}} \bibinfo{volume}{42}, \bibinfo{number}{4} (\bibinfo{year}{2020}).
\newblock


\bibitem[Metzler and Croft(2005)]%
        {Metzler2005AMR}
\bibfield{author}{\bibinfo{person}{Donald Metzler} {and} \bibinfo{person}{W.~Bruce Croft}.} \bibinfo{year}{2005}\natexlab{}.
\newblock \showarticletitle{A Markov random field model for term dependencies}. In \bibinfo{booktitle}{\emph{Annual International ACM SIGIR Conference on Research and Development in Information Retrieval}}.
\newblock


\bibitem[Mikolov et~al\mbox{.}(2013)]%
        {Mikolov2013EfficientEO}
\bibfield{author}{\bibinfo{person}{Tomas Mikolov} {et~al\mbox{.}}} \bibinfo{year}{2013}\natexlab{}.
\newblock \showarticletitle{Efficient Estimation of Word Representations in Vector Space}. In \bibinfo{booktitle}{\emph{ICLR}}.
\newblock


\bibitem[Mitra and Craswell(2018)]%
        {mitra2018an}
\bibfield{author}{\bibinfo{person}{Bhaskar Mitra} {and} \bibinfo{person}{Nick Craswell}.} \bibinfo{year}{2018}\natexlab{}.
\newblock \showarticletitle{An Introduction to Neural Information Retrieval}.
\newblock \bibinfo{journal}{\emph{Foundations and Trends® in Information Retrieval}} \bibinfo{volume}{13}, \bibinfo{number}{1} (\bibinfo{date}{December} \bibinfo{year}{2018}), \bibinfo{pages}{1--126}.
\newblock


\bibitem[Naseri et~al\mbox{.}(2021)]%
        {10.1007/978-3-030-72113-8_31}
\bibfield{author}{\bibinfo{person}{Shahrzad Naseri} {et~al\mbox{.}}} \bibinfo{year}{2021}\natexlab{}.
\newblock \showarticletitle{CEQE: Contextualized Embeddings for Query Expansion}. In \bibinfo{booktitle}{\emph{Advances in Information Retrieval: 43rd European Conference on IR Research, ECIR 2021.}} \bibinfo{publisher}{Springer-Verlag}, \bibinfo{address}{Berlin, Heidelberg}, \bibinfo{pages}{467–482}.
\newblock
\showISBNx{978-3-030-72112-1}


\bibitem[Neji et~al\mbox{.}(2021)]%
        {NEJI20211111}
\bibfield{author}{\bibinfo{person}{Sameh Neji} {et~al\mbox{.}}} \bibinfo{year}{2021}\natexlab{}.
\newblock \showarticletitle{HyRa: An Effective Hybrid Ranking Model}.
\newblock \bibinfo{journal}{\emph{Procedia Comput. Sci.}} \bibinfo{volume}{192}, \bibinfo{number}{C} (\bibinfo{date}{Jan} \bibinfo{year}{2021}), \bibinfo{pages}{1111–1120}.
\newblock
\showISSN{1877-0509}


\bibitem[Pang et~al\mbox{.}(2016)]%
        {https://doi.org/10.48550/arxiv.1606.04648}
\bibfield{author}{\bibinfo{person}{Liang Pang} {et~al\mbox{.}}} \bibinfo{year}{2016}\natexlab{}.
\newblock \bibinfo{title}{A Study of MatchPyramid Models on Ad-hoc Retrieval}.
\newblock
\newblock
\urldef\tempurl%
\url{https://doi.org/10.48550/ARXIV.1606.04648}
\showDOI{\tempurl}


\bibitem[Robertson et~al\mbox{.}(2004)]%
        {10.1145/1031171.1031181}
\bibfield{author}{\bibinfo{person}{Stephen Robertson} {et~al\mbox{.}}} \bibinfo{year}{2004}\natexlab{}.
\newblock \showarticletitle{Simple BM25 extension to multiple weighted fields}. In \bibinfo{booktitle}{\emph{Proceedings of the Thirteenth ACM International Conference on Information and Knowledge Management}} (Washington, D.C., USA) \emph{(\bibinfo{series}{CIKM '04})}. \bibinfo{pages}{42–49}.
\newblock
\showISBNx{1581138741}


\bibitem[Robertson and Zaragoza(2009)]%
        {Robertson2009ThePR}
\bibfield{author}{\bibinfo{person}{Stephen~E. Robertson} {and} \bibinfo{person}{Hugo Zaragoza}.} \bibinfo{year}{2009}\natexlab{}.
\newblock \showarticletitle{The Probabilistic Relevance Framework: BM25 and Beyond}.
\newblock \bibinfo{journal}{\emph{Found. Trends Inf. Retr.}}  \bibinfo{volume}{3} (\bibinfo{year}{2009}), \bibinfo{pages}{333--389}.
\newblock


\bibitem[Rocchio(1971)]%
        {rocchio71relevance}
\bibfield{author}{\bibinfo{person}{J.~J. Rocchio}.} \bibinfo{year}{1971}\natexlab{}.
\newblock \showarticletitle{Relevance feedback in information retrieval}.
\newblock In \bibinfo{booktitle}{\emph{The Smart retrieval system - experiments in automatic document processing}}, \bibfield{editor}{\bibinfo{person}{G.~Salton}} (Ed.). \bibinfo{publisher}{Englewood Cliffs, NJ: Prentice-Hall}, \bibinfo{pages}{313--323}.
\newblock


\bibitem[Shen et~al\mbox{.}(2014)]%
        {10.1145/2567948.2577348}
\bibfield{author}{\bibinfo{person}{Yelong Shen} {et~al\mbox{.}}} \bibinfo{year}{2014}\natexlab{}.
\newblock \showarticletitle{Learning semantic representations using convolutional neural networks for web search}. In \bibinfo{booktitle}{\emph{Proceedings of the 23rd International Conference on World Wide Web}} (Seoul, Korea) \emph{(\bibinfo{series}{WWW '14 Companion})}. \bibinfo{pages}{373–374}.
\newblock
\showISBNx{9781450327459}


\bibitem[Sivic and Zisserman(2003)]%
        {1238663}
\bibfield{author}{\bibinfo{person}{Sivic} {and} \bibinfo{person}{Zisserman}.} \bibinfo{year}{2003}\natexlab{}.
\newblock \showarticletitle{Video Google: a text retrieval approach to object matching in videos}. In \bibinfo{booktitle}{\emph{Proceedings Ninth IEEE International Conference on Computer Vision}}. \bibinfo{pages}{1470--1477 vol.2}.
\newblock


\bibitem[Soboroff(2021)]%
        {soboroff2021overview}
\bibfield{author}{\bibinfo{person}{Ian Soboroff}.} \bibinfo{year}{2021}\natexlab{}.
\newblock \showarticletitle{Overview of TREC 2021}. In \bibinfo{booktitle}{\emph{30th Text REtrieval Conference. Gaithersburg, Maryland}}.
\newblock


\bibitem[Svore and Burges(2009)]%
        {10.1145/1645953.1646237}
\bibfield{author}{\bibinfo{person}{Krysta~M. Svore} {and} \bibinfo{person}{Christopher~J.C. Burges}.} \bibinfo{year}{2009}\natexlab{}.
\newblock \showarticletitle{A machine learning approach for improved BM25 retrieval}. In \bibinfo{booktitle}{\emph{Proceedings of the 18th ACM Conference on Information and Knowledge Management}} (Hong Kong, China) \emph{(\bibinfo{series}{CIKM '09})}. \bibinfo{numpages}{4}~pages.
\newblock
\showISBNx{9781605585123}


\bibitem[Turney and Pantel(2010)]%
        {10.5555/1861751.1861756}
\bibfield{author}{\bibinfo{person}{Peter~D. Turney} {and} \bibinfo{person}{Patrick Pantel}.} \bibinfo{year}{2010}\natexlab{}.
\newblock \showarticletitle{From Frequency to Meaning: Vector Space Models of Semantics}.
\newblock \bibinfo{journal}{\emph{J. Artif. Int. Res.}} \bibinfo{volume}{37}, \bibinfo{number}{1} (\bibinfo{date}{jan} \bibinfo{year}{2010}), \bibinfo{pages}{141–188}.
\newblock
\showISSN{1076-9757}


\bibitem[Vaswani et~al\mbox{.}(2017)]%
        {NIPS2017_3f5ee243}
\bibfield{author}{\bibinfo{person}{Ashish Vaswani} {et~al\mbox{.}}} \bibinfo{year}{2017}\natexlab{}.
\newblock \showarticletitle{Attention is All you Need}. In \bibinfo{booktitle}{\emph{Advances in Neural Information Processing Systems}}, Vol.~\bibinfo{volume}{30}.
\newblock


\bibitem[Voorhees et~al\mbox{.}(2021)]%
        {10.1145/3451964.3451965}
\bibfield{author}{\bibinfo{person}{Ellen Voorhees} {et~al\mbox{.}}} \bibinfo{year}{2021}\natexlab{}.
\newblock \showarticletitle{TREC-COVID: Constructing a Pandemic Information Retrieval Test Collection}.
\newblock \bibinfo{journal}{\emph{SIGIR Forum}} \bibinfo{volume}{54}, \bibinfo{number}{1}, Article \bibinfo{articleno}{1} (\bibinfo{date}{feb} \bibinfo{year}{2021}), \bibinfo{numpages}{12}~pages.
\newblock
\showISSN{0163-5840}


\bibitem[Wang et~al\mbox{.}(2020)]%
        {wang-etal-2020-cord}
\bibfield{author}{\bibinfo{person}{Lucy~Lu Wang} {et~al\mbox{.}}} \bibinfo{year}{2020}\natexlab{}.
\newblock \showarticletitle{{CORD-19}: The {COVID-19} Open Research Dataset}. In \bibinfo{booktitle}{\emph{Proceedings of the 1st Workshop on {NLP} for {COVID-19} at {ACL} 2020}}. \bibinfo{address}{Online}.
\newblock


\bibitem[Wang et~al\mbox{.}(2023)]%
        {10.1145/3572405}
\bibfield{author}{\bibinfo{person}{Xiao Wang} {et~al\mbox{.}}} \bibinfo{year}{2023}\natexlab{}.
\newblock \showarticletitle{ColBERT-PRF: Semantic Pseudo-Relevance Feedback for Dense Passage and Document Retrieval}.
\newblock \bibinfo{journal}{\emph{ACM Trans. Web}} \bibinfo{volume}{17}, \bibinfo{number}{1}, Article \bibinfo{articleno}{3} (\bibinfo{date}{jan} \bibinfo{year}{2023}), \bibinfo{numpages}{39}~pages.
\newblock
\showISSN{1559-1131}


\bibitem[Williams et~al\mbox{.}(2014)]%
        {10.1145/2644866.2644895}
\bibfield{author}{\bibinfo{person}{Kyle Williams} {et~al\mbox{.}}} \bibinfo{year}{2014}\natexlab{}.
\newblock \showarticletitle{SimSeerX: a similar document search engine}. In \bibinfo{booktitle}{\emph{Proceedings of the 2014 ACM Symposium on Document Engineering}} (Fort Collins, Colorado, USA) \emph{(\bibinfo{series}{DocEng '14})}. \bibinfo{pages}{143–146}.
\newblock
\showISBNx{9781450329491}


\bibitem[Xiong et~al\mbox{.}(2020)]%
        {Xiong2020ApproximateNN}
\bibfield{author}{\bibinfo{person}{Lee Xiong} {et~al\mbox{.}}} \bibinfo{year}{2020}\natexlab{}.
\newblock \showarticletitle{Approximate Nearest Neighbor Negative Contrastive Learning for Dense Text Retrieval}. In \bibinfo{booktitle}{\emph{ICLR}}.
\newblock


\bibitem[Yu et~al\mbox{.}(2021)]%
        {10.1145/3459637.3482124}
\bibfield{author}{\bibinfo{person}{HongChien Yu} {et~al\mbox{.}}} \bibinfo{year}{2021}\natexlab{}.
\newblock \showarticletitle{Improving Query Representations for Dense Retrieval with Pseudo Relevance Feedback}. In \bibinfo{booktitle}{\emph{Proceedings of the 30th ACM International Conference on Information \& Knowledge Management}} (Virtual Event, Queensland, Australia) \emph{(\bibinfo{series}{CIKM '21})}. \bibinfo{pages}{3592–3596}.
\newblock
\showISBNx{9781450384469}


\bibitem[Yuan et~al\mbox{.}(2012)]%
        {DBLP:conf/eccv/YuanGCLJ12}
\bibfield{author}{\bibinfo{person}{Jiangbo Yuan} {et~al\mbox{.}}} \bibinfo{year}{2012}\natexlab{}.
\newblock \showarticletitle{Efficient Mining of Repetitions in Large-Scale {TV} Streams with Product Quantization Hashing}. In \bibinfo{booktitle}{\emph{Computer Vision - {ECCV} 2012. Workshops and Demonstrations - Florence, Italy.}}, Vol.~\bibinfo{volume}{7583}. \bibinfo{publisher}{Springer}, \bibinfo{pages}{271--280}.
\newblock


\bibitem[Zhai and Lafferty(2001)]%
        {10.1145/502585.502654}
\bibfield{author}{\bibinfo{person}{Chengxiang Zhai} {and} \bibinfo{person}{John Lafferty}.} \bibinfo{year}{2001}\natexlab{}.
\newblock \showarticletitle{Model-based feedback in the language modeling approach to information retrieval}. In \bibinfo{booktitle}{\emph{Proceedings of the 10th International Conference on Information and Knowledge Management}} \emph{(\bibinfo{series}{CIKM '01})}.
\newblock
\showISBNx{1581134363}


\end{thebibliography}


\end{document}